\newif\ifjcs\jcstrue
\ifjcs
\documentclass[a4paper]{article}
\else
\documentclass[10pt, conference, compsocconf]{IEEEtran}
\fi
 
\usepackage[utf8]{inputenc}
\usepackage{xspace}
\usepackage{amsmath}
\usepackage{amssymb}
\usepackage{xcolor}
\ifjcs
  \usepackage{fmtcount}
  \usepackage{amsthm}
  \usepackage{multicol}
\fi

\usepackage{listings}
\usepackage{versions}
\usepackage{graphicx}   
\usepackage[
colorlinks=true,
linkcolor=linkColor,
citecolor=linkColor,
urlcolor=linkColor,
pdftitle={Guiding a General-Purpose C Verifier to Prove Cryptographic Protocols},
pdfsubject={},
pdfauthor={Dupressoir,Gordon,Juerjens,Naumann},
pdfkeywords={}
]{hyperref}

\usepackage{coqdoc} 

\newcommand{\correction}[1]{{\color{blue}#1}}
\renewcommand{\correction}[1]{#1} 
  
\newif\iffull\fulltrue  
\newif\ifdraft\drafttrue 

\draftfalse\fulltrue

\ifdraft
\def\Blue{\color{blue}}
\def\Black{\color{black}}

\newcommand{\displaycomment}[1]{ #1 {}}
\includeversion{DRAFT}
\else
\def\Blue{}\def\Black{}
\newcommand{\displaycomment}[1]{}
\excludeversion{DRAFT}
\fi

\iffull
\newenvironment{FULL}{\ifdraft\Blue\fi}{\ifdraft\Black\fi}
\excludeversion{SHORT}
\else
\excludeversion{FULL}
\includeversion{SHORT}
\fi

\ifjcs
\includeversion{JCS}{}{}
\excludeversion{IEEE}
\else
\excludeversion{JCS}
\includeversion{IEEE}
\fi

\iffull\begin{IEEE}\usepackage{msrtr-cam}\end{IEEE}\fi

\definecolor{dkblue}{rgb}{0,0.1,0.5}
\definecolor{dkgreen}{rgb}{0,0.4,0}
\definecolor{dkred}{rgb}{0.6,0,0}
\definecolor{linkColor}{rgb}{0,0.1,0.5}

\newcommand{\todonote}[2]{
\displaycomment{{\color{dkred}{\bf#1: }#2}}%
\typeout{#1 TODO: (\thepage) #2}}
\newcommand{\Remark}[2]{
\displaycomment{{{\color{dkgreen}(#1: #2)}}}}

\newcommand{\adg}[1]{\Remark{Andy}{#1}}
\newcommand{\FD}[1]{\todonote{Fran{\c{c}}ois}{#1}}
\newcommand{\fd}[1]{\Remark{Fran{\c{c}}ois}{#1}}
\newcommand{\DN}[1]{\todonote{Dave}{#1}}
\newcommand{\dn}[1]{\Remark{Dave}{#1}}

\newcommand{\jj}[1]{\Remark{Jan}{#1}}

\newcommand{\hbra}{
  \hbox to \columnwidth{\vrule width0.3mm height 1.8mm depth-0.3mm
    \leaders\hrule height1.8mm depth-1.5mm\hfill
    \vrule width0.3mm height 1.8mm depth-0.3mm}}
\newcommand{\hket}{
  \hbox to \columnwidth{\vrule width0.3mm height1.5mm
    \leaders\hrule height0.3mm\hfill
    \vrule width0.3mm height1.5mm}}

\makeatletter
  \newcommand{\addToLabel}[1]{%
    \protected@edef\@currentlabel{\@currentlabel#1}%
  }
\makeatother
\newcommand{\ratio}{.35}

\newenvironment{display}[2][\ratio]{\vspace{-0.5ex}
\par\begin{tabbing}
    \hspace{1.5em} \= \hspace{#1\columnwidth-1.5em} \= \hspace{1.5em} \= \kill
    \textbf{\normalsize #2}\\[-.8ex]
    \hbra\\[-.8ex]
  }{\\[-.8ex]\hket
  \end{tabbing}\vspace{-1.5ex}}

\newcommand{\vccdisplayendspace}{-3ex}

\newenvironment{vccdisplay}[2][-3]{\renewcommand{\vccdisplayendspace}{#1ex}
  \vspace{0.5ex}\par
  \noindent\textbf{\normalsize #2}\\[-1ex]
  \hbra\vspace{-4.5ex}
}{\vspace{\vccdisplayendspace}\hket}

\newcommand{\entry}[2]{\>$#1$\>\>#2}
\newcommand{\clause}[2]{$#1$\>\>#2}
\newcommand{\Category}[2]{\clause{#1::=}{#2}}

\newcounter{rule}

\newcommand{\staterule}[4][]{%
  \refstepcounter{rule}%
  \addToLabel{(#2)}\label{#2}%
  $\begin{array}[b]{@{}l@{}}%
    \mbox{(#2)#1}\\%
    \begin{array}{@{}c@{}}
      #3\\
      \hline
      \raisebox{0ex}[2.5ex]{\strut}#4%
    \end{array}
  \end{array}$}

\newtheorem{assumption}{Assumption}
\newtheorem{theorem}{Theorem}
\newtheorem{lemma}{Lemma}
\newtheorem{definition}{Definition}

\lstdefinelanguage{VCC}[ANSI]{C}{%
  morekeywords={axiom,theorem,rule,
                              requires,ensures,maintains,always,
                              reads,writes,
                              Hint,assert,assume,
                              invariant,
                              wrap,unwrap,atomic,unchecked,spec,
                              bool,NULL,true,false,
                              mutable,thread_local,wrapped,unwrapped,
                              old,unchanged,span,emb,
                              ispure,frameaxiom,
		         result,this,forall,exists,lambda,owns,
                              def,abstract,datatype,record},
  columns = fullflexible,
  moredelim=[is][\fontfamily{cmtt}\selectfont\textcolor{vccOutColor}]{/*`}{`*/},
  morecomment=[is]{//--}{//--},
  basicstyle=\scriptsize\fontfamily{cmr},
  breaklines=true,
  rangebeginprefix=//Begin\ ,
  rangeendprefix=//End\ ,
  rangesuffix=,
  includerangemarker=false,
  literate= {int\ dummy;}{\ }1
  {==>}{{$\Rightarrow$}}1
  {\\forall}{{$\forall{}$}}1
  {\\exists}{{$\exists{}$}}1
  {\\lambda}{{$\lambda{}$}}1,
}

\lstdefinelanguage{coq}%
  {morekeywords={Variable,Inductive,CoInductive,Fixpoint,CoFixpoint,
      Definition,Lemma,Theorem,Axiom,Local,Save,Grammar,Syntax,Intro,
      Trivial,Qed,Intros,Symmetry,Simpl,Rewrite,Apply,Elim,Assumption,
      Left,Cut,Case,Auto,Unfold,Exact,Right,Hypothesis,Pattern,Destruct,
      Constructor,Defined,Fix,Record,Proof,Induction,Hints,Exists,let,in,
      Parameter,Split,Red,Reflexivity,Transitivity,if,then,else,Opaque,
      Transparent,Inversion,Absurd,Generalize,Mutual,Cases,of,end,Analyze,
      AutoRewrite,Functional,Scheme,params,Refine,using,Discriminate,Try,
      Require,Load,Import,Scope,Set,Open,Section,End,match,with,Ltac 
	}, 
   columns = fullflexible, 
   basicstyle={\scriptsize\sffamily},
   moredelim=[is][\fontfamily{cmtt}]{(*`}{`*)},
   breaklines=true,
   literate={=>}{{$\Rightarrow$}}1 {>->}{{$\rightarrowtail$}}2{->}{{$\to$}}1
   {\/\\}{{$\wedge$}}1
   {|-}{{$\vdash$}}1
   {\\\/}{{$\vee$}}1
   {~}{{$\sim$}}1
   {forall}{{$\forall$}}1
   {exists}{{$\exists$}}1
  }
\newcommand{\coqkw}{\lstinline[language={coq},basicstyle={\normalsize\sffamily}]}

\newcommand{\iface}{\mathcal{I}}
\newcommand{\good}[4]{#1;#2 \vdash #3\mbox{\textvisiblespace}\ #4}
\newcommand{\kw}[1]{\mbox{\lstinline[language={VCC},basicstyle={\footnotesize\sffamily}]{#1}}}
\renewcommand{\log}{\mathcal{L}}
\newcommand{\kwCprime}{\kw{C}\ensuremath{'}}
\ifjcs\newenvironment{IEEEproof}{\begin{proof}}{\end{proof}}\fi

\newenvironment{prog}{\begin{array}[t]{@{}l@{}}}{\end{array}}

\title{Guiding a General-Purpose C Verifier to Prove Cryptographic Protocols}

\ifjcs
\author{Fran{\c{c}}ois~Dupressoir
\and Andrew~D.~Gordon
\and Jan~J{\"u}rjens
\and David~A.~Naumann}
\else
\author{
\IEEEauthorblockN{Fran{\c{c}}ois~Dupressoir}
\IEEEauthorblockA{The Open University}
\and
\IEEEauthorblockN{Andrew~D.~Gordon}
\IEEEauthorblockA{Microsoft Research}
\and
\IEEEauthorblockN{Jan~J{\"{u}}rjens}
\IEEEauthorblockA{TU Dortmund \& Fraunhofer ISST}
\and
\IEEEauthorblockN{David~A.~Naumann}
\IEEEauthorblockA{Stevens Institute of Technology}
}
\fi
\date{}

\begin{document}
\begin{FULL}
\ifjcs
\else
\begin{DRAFT}
This is technical report MSR--TR--2011--50.
\tableofcontents
\clearpage
\end{DRAFT}
\date{May 2011}
\msrtrconf{An abridged form of this report appears in the proceedings of the
  \emph{24th IEEE Computer Security Foundations Symposium}, June 27--29, 2011,
  Domaine de l'Abbaye des Vaux de Cernay, France.}
\msrtrno{MSR--TR--2011--50}
\msrtrmaketitle
\thispagestyle{empty}
\setcounter{page}{2}
\fi
\end{FULL}

\maketitle
\iffull\pagestyle{plain}\fi

\begin{abstract}
We describe how to verify security properties of C code for cryptographic protocols by using a general-purpose verifier.
We prove security theorems in the symbolic model of cryptography.
Our techniques include:
use of ghost state to attach formal algebraic terms to concrete byte arrays
and to detect collisions when two distinct terms map to the same byte array;
decoration of a crypto API with contracts based on symbolic terms;
and expression of the attacker model in terms of C programs.
We rely on the general-purpose verifier VCC;
we guide VCC to prove security simply by writing suitable header files and annotations in implementation files,
rather than by changing VCC itself.
We formalize the symbolic model in Coq in order to justify the addition of axioms to VCC.
\end{abstract}

\ifjcs\else
\IEEEpeerreviewmaketitle
\fi

\FD{Check line breaks in imported code before sending out for publication}
\dn{Don't spend too much time on line breaks; even using the publisher's macros, the final version is often formatted more compactly.}
\FD{Keep references clean and tidy. Make sure full info is given.}

\section{Introduction}\label{sec:intro}

\adg{Policy: catenation instead of concatenation, to save trees}
Economies of scale suggest that it is better,
where possible, to adapt an existing general-purpose tool to a specialist problem,
than to go to the expense of building a specialist tool for each niche application area.

Our particular concern is the specialist problem of verifying the implementation code of cryptographic protocols
\cite{goubault-larrecq_cryptographic_2005,DBLP:conf/lics/Gordon06,BFGT08:VerifiedInteroperableImplementationsOfSecurityProtocols}.
This code is mostly written in C, and is often the first---and sometimes the only---completely precise description of the message formats and invariants of cryptographic protocols.
Hence, reasoning about such code offers a way to find and prevent both the design and implementation flaws that lead to expensive failures
(for instance, \cite{DBLP:conf/asian/CervesatoJSTW06,ocert_ocert_2009,DBLP:conf/sp/MurdochDAB10}).

\paragraph{Background: Proving Cryptographic Protocol Code}


The prior work on verifying C code of security protocols relies on special-purpose tools.
Csur~\cite{goubault-larrecq_cryptographic_2005} analyzes C code for secrecy properties via a custom abstract interpretation,
while ASPIER~\cite{chaki_aspier:automated_2008} relies on security-specific software model-checking techniques,
obtaining good results on the main loop of OpenSSL.
Both these tools use the symbolic model of cryptography introduced by Dolev and Yao~\cite{dolev_security_1983}.

\begin{JCS}
More recently, symbolic execution was used to extract partial models from C programs, translating them into ProVerif processes that can be verified~\cite{AGJ11}.
For some of these models, computational soundness results can be used to obtain security theorems in the computational model of cryptography for a single path of the C program, identified by dynamic execution.
In subsequent work \cite{AGJ12}, the translation and model are adapted slightly, to produce CryptoVerif models for direct verification in the computational model.
However, the limitation to a single path is still present.
\end{JCS}

Another line of work considers the problem of verifying reference implementations written as functional programs.
Initial approaches rely on security-specific analyzers.
%
%
The tools FS2PV~\cite{BFGT08:VerifiedInteroperableImplementationsOfSecurityProtocols} and FS2CV~\cite{fs2cv} translate functional programs in {F\#} to the process calculi
accepted by the specialised verifiers ProVerif~\cite{DBLP:conf/csfw/Blanchet01} and CryptoVerif~\cite{DBLP:conf/sp/Blanchet06}
for automatic verification in the symbolic and computational models;
an implementation of TLS~\cite{bhargavan_cryptographically_2008} is a substantial case study.

Instead of translating to a protocol verifier,
the typechecker F7~\cite{DBLP:journals/toplas/BengtsonBFGM11} checks {F\#} by using security-specific refinement types, types qualified with formulas,
to express security properties.
The theory of F7 is based on the symbolic model,
although in some circumstances F7 can be adapted to the computational model~\cite{BMU10:CompSVSC,Fournet11}.

A subsequent, more flexible, method of using refinement types, based on \emph{invariants for cryptographic structures}~\cite{bhargavan_modular_2010},
relies on axiomatizations of cryptographic predicates (such as which data is public);
first implemented for F7, the method works in principle with any general-purpose refinement-type checker.
\begin{FULL}%
The method has been ported to F*~\cite{SCFBY10:SDPVDT}, a recent functional language.%
\end{FULL}

Our strategy is to port this method to a verifier for C.
 
\paragraph{Background: General-Purpose C Verifiers}

By now there are several general-purpose and more-or-less automatic verification tools for C,
including Frama-C~\cite{Frama-C},
VeriFast~\cite{JP09:VeriFast},
and VCC~\cite{cohen_vcc:practical_2009}.
This paper describes our techniques for guiding one of these, VCC, to verify a range of security protocol implementations.
Although we adopt VCC, we expect our method would port to other tools.

VCC verifies C code against specifications written as
\begin{IEEE}function contracts in the tradition of Floyd-Hoare logic.\end{IEEE}
\begin{JCS} conventional function contracts, that is, pre, post, and frame conditions.\end{JCS}
It translates C to an intermediate verifier, Boogie~\cite{DBLP:conf/fmco/BarnettCDJL05}, which itself relies on an SMT solver, Z3~\cite{DBLP:conf/tacas/MouraB08}.
The translation to Boogie encodes an accurate low-level memory model for C.
VCC supports concurrency, which we use to model distributed execution of protocols as well as for multithreaded code.
Specifications use  \emph{ghost state}, that is, specially-marked program variables that may be mentioned in contracts, but that are not allowed to affect ordinary state or control flow.
We aim to scale to verify large amounts of off-the-shelf C code such as OpenSSL;
VCC has already proved itself capable of verifying large pre-existing codebases \cite{cohen_vcc:practical_2009}.


\subsection{Outline of our Techniques}

We summarize the main aspects of our adaptation to C and VCC of the method of \emph{invariants for cryptographic structures} \cite{bhargavan_modular_2010}.
Later on, we describe the differences from the prior work on {F\#} and F7.

\subsubsection{Language-independent definitional theory}
We develop a theory of symbolic cryptography, independent of any programming language, within the interactive proof assistant Coq.
The theory is definitional in the sense that it is developed from sound definitional principles (on the basis of symbolic cryptography), with no
additional assumptions.

As usual in the symbolic model~\cite{dolev_security_1983}, the core of our theory is an algebraic type with constructors corresponding to the following:
the outcomes of cryptographic algorithms such as keyed hashing, encryptions,  and signatures;
literal byte strings (which represent messages, principal names, keys, and nonces, etc);
and reversible pairing (used for message formatting and implemented with a length field).

Our theory accounts for the time-dependent history of protocol execution by defining a \emph{log}, $\log$, to be a set of \emph{events}, which records progress so far in a protocol run.
\begin{FULL}%
Events are values of another algebraic type, for example:
\begin{itemize}
\item
Key generation is recorded by an event such as $\textbf{New}~k~(\textbf{KeyAB}~a~b)$,
which records that the term $k$ represents a fresh key with usage $\textbf{KeyAB}~a~b$,
that is, a key shared between principals $a$ and $b$.
Values of the algebraic datatype $\kw{usage}$ represent the different sorts of keys used in a protocol.
\item
Progress through a protocol is tracked by events such as $\textbf{Request}~a~b~t$,
meaning that a client $a$ has started a protocol run by sending a request $t$ to a server $b$.
\item
Principal compromise is recorded using an event such as $\textbf{Bad}~a$,
meaning that principal $a$, and its keys, are under the control of the attacker.
\end{itemize}%
\end{FULL}

Our theory also includes an inductive definition of confidentiality levels of terms, parameterized by the log of events.
\begin{IEEE}
Terms may either be public (known to the attacker), or private (known only to the protocol participants).
\end{IEEE}
\begin{JCS}
We borrow the \coqkw{High}/\coqkw{Low} terminology from the information-flow literature,
although we give them a related, but different, meaning, reminiscent of Abadi's \textit{Any}/\textit{Public} security levels~\cite{Abadi:1999:STS:324133.324266}.
All terms manipulated during a protocol run must be \coqkw{High},
and we verify that all terms that become known to the adversary are \coqkw{Low}.
By definition, every \coqkw{Low} term is also \coqkw{High}.
A term is said to be secret if it is not \coqkw{Low}.
\end{JCS}
We need to parameterize by the log because once events such as a principal compromise are logged, more data becomes public (such as keys known to the principal).

\subsubsection{Theory imported as first-order axioms}
C is a low-level language and does not directly support  abstractions for algebraic types.
\begin{IEEE}
Also, VCC cannot easily perform proofs by induction, being based on first-order logic without induction principles.
%
%
Still, VCC does allow ghost state and ghost commands to manipulate unbounded data---ghost data of type
\kw{mathint} consists of arbitrary mathematical integers---and VCC allows us to assume arbitrary first-order axioms.
Hence, we can import our Coq theory into VCC by (1) using ghost data of type \kw{mathint} to represent algebraic types such as symbolic terms,
and (2) importing Coq theorems about our inductive definitions as first-order axioms.
\end{IEEE}
\begin{JCS}
Also, VCC cannot easily perform the more complex proofs by induction, being based on first-order logic and having only limited support for induction principles.
Moreover, performing the security proof in Coq allows us, in theory, to re-use a protocol description and proof of security for different implementations, written in different languages or verified using different tools.
We therefore use VCC inductive datatypes and functions to define
the types of terms, key usages, events and logs, as well as basic predicates
on these types, but still import the definition of the \coqkw{Level} predicate
and related theorems as first-order axioms.
The adversary only directly manipulates concrete C data,
and cannot (for example) pattern match encryption terms to extract the key or payload.
\end{JCS}
Results proved by VCC hold in all models of the axioms, including the intended one inductively defined as the \coqkw{Level} predicate in Coq.

\subsubsection{A ghost table relates bytestrings and symbolic terms}
To simplify the example, our cryptographic library manipulates byte arrays via a C struct \kw{bytes_c}, which contains a length field together with a pointer to a heap-allocated chunk of memory with that length.
Additionally, the struct contains a
\begin{IEEE}\kw{mathint}\end{IEEE}
ghost field \kw{encoding},
\begin{JCS}that has the type \kw{bytes} of mathematical bytestring values,\end{JCS}
satisfying an invariant that it encodes the actual bytestring stored in memory.
\begin{IEEE}
In global ghost state, we maintain
a \emph{representation table}, which holds a finite one-to-one correspondence between bytestrings and symbolic terms.
These are the cryptographically significant bytestrings arising so far in the run.
The predicate $\kw{table.DefinedB}[b]$ holds just if bytestring $b$ exists in the table.
If so, $\kw{table.B2T}[b]$ is the corresponding term.
Conversely, if term $t$ is in the table, $\kw{table.T2B}[t]$ is the corresponding bytestring.
\end{IEEE}
\begin{JCS}
We maintain a global ghost variable \kw{CS}, which represents the cryptographic state,
composed of both the log, and a \emph{representation table}.
The latter holds a finite one-to-one correspondence betweeen symbolic terms and
the cryptographically significant bytestrings arising so far in the run.
The predicate \kw{CS.T.DefinedB[b]} holds just if bytestring \kw{b} exists in the table.
If so, \kw{CS.T.B2T[b]} is the corresponding term.
Conversely, if term \kw{t} is in the table, \kw{CS.T.T2B[t]} is the corresponding bytestring.
\end{JCS}

We rely on the table to write VCC function contracts that specify symbolic assumptions about concrete cryptographic routines.
For example, the contract for \kw{hmacsha1} follows.
It enforces that $t_b$ can be MAC'ed with $t_k$ only when
the protocol-dependent precondition $\kw{canHmac}(t_k,t_b)$ holds in the current log \kw{CS.L}.
\begin{IEEE}
\begin{lstlisting}[language=VCC]
int hmacsha1(bytes_c *k, bytes_c *b, bytes_c *res)
requires(table.DefinedB[k->encoding])
requires(table.DefinedB[b->encoding])
ensures(!result ==> table.DefinedB[res->encoding])
requires
  (MACSays(table.B2T[k->encoding],table.B2T[b->encoding]))
ensures(!result ==>
  table.B2T[res->encoding] ==
    Hmac(table.B2T[k->encoding],table.B2T[b->encoding]));
\end{lstlisting}
\end{IEEE}
\begin{JCS}
\begin{lstlisting}[language=VCC]
int hmacsha1(bytes_c *k, bytes_c *b, bytes_c *res)
_(requires CS.T.DefinedB[k->encoding])
_(requires CS.T.DefinedB[b->encoding])
_(ensures !\result ==> CS.T.DefinedB[res->encoding])
_(requires
   canHmac(CS.T.B2T[k->encoding],CS.T.B2T[b->encoding],CS.L))
_(ensures !\result ==>
  CS.T.B2T[res->encoding] ==
    Hmac(CS.T.B2T[k->encoding],CS.T.B2T[b->encoding]));
\end{lstlisting}
\end{JCS}

The contract's first three lines express the precondition (using the \kw{requires} keyword) that the two concrete arguments are in the table, and the postcondition (using the \kw{ensures} keyword) that the concrete result is in the table.
The fourth line requires \correction{that} the $\kw{canHmac}$ predicate is fulfilled \correction{by the input terms in the current log}.
The final line ensures \correction{that} the term associated with the result is $\kw{Hmac}(t_k,t_b)$, where terms $t_k$ and $t_b$ are associated with the concrete inputs.
\begin{IEEE}
(As these lines illustrate, we include the ghost field $\kw{encoding}$ in $\kw{bytes_c}$ because it allows succinct access in specifications to the contents of memory.)
\end{IEEE}
\begin{JCS}
(As this contract illustrates, the inclusion of ghost field $\kw{encoding}$ in $\kw{bytes_c}$ allows succinct access in specifications to the contents of memory.)
\end{JCS}

The VCC-verified concrete implementation of \kw{hmacsha1}, called a \emph{hybrid wrapper},
simply calls a routine trusted to compute the MAC of the inputs,
and then in ghost code updates the table with the result, if it is new.

\subsubsection{Protocol roles described as ordinary C code}
Each role of a protocol is simply code in C, executed as normal, and verified for memory safety and security with VCC.
We model distributed execution by multiple threads that communicate concretely by message passing via a network API, but that share a single
\begin{IEEE}representation table.\end{IEEE}
\begin{JCS}cryptographic state.\end{JCS}
The protocol code can itself be multithreaded and use shared memory, but that feature is not used in the simple examples presented here.

Throughout this article, we take as a running example the following simple authenticated RPC \correction{protocol (introduced in} \cite{bhargavan_modular_2010}).
This two-party protocol uses a pre-shared secret key to authenticate requests and responses and link responses to the corresponding request, using a keyed hash as MAC.
\begin{SHORT}Our verified C code for this protocol is available in the full version online.\end{SHORT}

\begin{small}
\begin{display}{Running example: an authenticated RPC protocol}
a $\rightarrow$ B\=\kill
a                      \>: \kw{Log(Request(a, b, payload))}\\
a $\rightarrow$ b: payload $\vert$ hmac(kab, "1" $\vert$ payload)\\
b		\>: \kw{assert(Request(a, b, payload))}\\
b                      \>: \kw{Log(Response(a, b, payload, payload'))}\\
b $\rightarrow$ a: payload' $\vert$ hmac(kab, "2" $\vert$ payload $\vert$ payload')\\
a                      \>: \kw{assert(Response(a, b, payload, payload'))}
\end{display}
\end{small}

The narration logs events marking $a$'s intent to send a request, and $b$'s intent to send a response.
At these points in our code, ghost commands add events to the log.
The narration also includes correspondence assertions marking $b$'s conclusion that $a$ has sent a request,
and $a$'s conclusion that $b$ has sent a response.
In our code, these correspondences become $\kw{assert}$ statements, to be verified by VCC.

\subsubsection{Attacker model expressed using C interface}
We prove protocol code secure against a network-based attacker~\cite{needham_using_1978},
rather than against say local malware.
We consider an attacker to be a C program consisting simply of a series of calls to functions in the \emph{attacker interface}.
In keeping with the symbolic model, the attacker cannot directly manipulate bytestrings using the bitwise operators of C,
but only via this interface.
It includes functions for cryptography, to send and receive network messages, to create new principals (but without access to their keys),
to create instances of protocol roles, and to cause the compromise of principals (after which their keys are available).\adg{maybe cite TulaFale or applied $\pi$ work that introduced this}
The addition of explicit compromise allows us to reason about properties that may hold even after some keys become public.
For example, when multiple parallel sessions are concerned, it is desirable that a compromise in one of them does not affect security properties in the others.
Since our security results hold in spite of an arbitrary attacker, we place no bound on the number of distinct principals or concurrent sessions.

\subsubsection{Security theorems obtained via a general-purpose verifier}
By running VCC on the protocol implementation,
we prove both correspondence properties, expressing authentication and integrity properties,
and weak secrecy properties.
As mentioned, correspondences amount to embedded \kw{assert} statements.
Standard symbolic cryptography assumptions are expressed using local \kw{assume} statements.
Weak secrecy properties amount to consequences of invariants, respected by all verified code.
A typical secrecy property can be explained as follows:
if $k$ is a key shared between $a$ and $b$ and $k$ is public,
then either principal $a$ or principal $b$ is compromised.
Proof with VCC is semi-automatic in that it relies on automatic deductive inference, but with the help of user-supplied annotations.

\subsection{Contributions of the Paper}

To the best of our knowledge, this is the first published verification methodology for C implementations of cryptographic protocols
that proves both memory safety and security properties for unbounded sessions.
Csur~\cite{goubault-larrecq_cryptographic_2005\iffull,goubault-larrecq_cryptographic_2009\fi}
proves secrecy properties, but does not show memory safety;
in fact, verification succeeds despite the example code allowing accidental access to uninitialized memory.
ASPIER~\cite{chaki_aspier:automated_2008} proves various security properties by
software model-checking, but verification considers only a few concurrent sessions,
and relies on substantial abstractions.

Our work takes the idea of invariants for cryptographic structures~\cite{bhargavan_modular_2010}
away from strongly-typed functional programming in {F\#} and F7,
and recasts it in the setting of a weakly-typed low-level imperative language.
In C we can neither rely on abstract types nor escape from the difficulties of reasoning about mutable memory and aliasing.
These are probably the main new difficulties we address compared to the prior work with F7~\cite{bhargavan_modular_2010}; verifying C is much harder than verifying {F\#}.
In return, we enjoy vastly wider applicability, as the bulk of production cryptographic protocols is in C.
A less obvious and more technical benefit is that in F7,
the log of events is implicit and its impact on inductively defined predicates requires a bespoke notion of semantic safety for {F\#}.
By making the log explicit in ghost state,
we can work within a completely standard semantics of assertions on C programs.

Our method forces the developer to precisely specify memory safety and security properties. 
We verify them with a scalable and practically reliable tool that has clear semantics in terms of standard C, its compilers and hardware architecture.
Since user interaction is by way of code annotation, the verification effort may be expected to evolve well as the code base evolves.
We may also hope to reap the benefit of ongoing improvements in automation for general-purpose verifiers for C.
\begin{JCS}
These improvements have already had an impact while carrying out the work reported in this paper.
Benefiting from such improvements may mean that existing annotations need to be updated to respond to changes in the verifier used; however, in our experience this investment is worthwhile in terms of
the performance improvements gained (in our case, the adjustments necessitated by switching to a new version of VCC required about one person day effort).
\end{JCS}
Although we prove our security-specific theory in Coq,
we do trust the VCC/Boogie/Z3 tool chain and the C compiler.
%
%

\begin{IEEE}
We have validated our approach on implementations we developed using our own cryptographic APIs.
In future work, we intend to apply our techniques to pre-existing code.
\end{IEEE}
\begin{JCS}
We have validated our approach on implementations we developed, some using our own cryptographic APIs and with easy verification in mind~\cite{DGJN11:Guiding}, and some written in a more idiomatic C style involving manipulation of raw pointers to bytestrings, inlined marshalling and unmarshalling of pairs, and using existing cryptographic and network APIs~\cite{FAST11:Csec}.
In subsequent work~\cite{PM12}, VCC was used to verify security properties of existing code.
The approach used is somewhat different, as the entire proof is done in VCC, using a destructor-based definition of the \coqkw{Level} predicate (inspired by TAPS~\cite{cohen_first-order_2003}) and a set of attacker capabilities that is explicitly specified as a set of functions.

We improve on our previous paper by unifying the \coqkw{Pub} and \coqkw{Bytes} predicates into a single, unified, \coqkw{Level} predicate, which allows a much more general form of cryptographic invariants and usage predicates.
We also modify the implementation of those invariants and predicates in VCC
to leverage both recent changes to the verifier and lessons learned whilst using it,
significantly improving the performance of our method by letting the automated tool optimize the axioms,
and also improving the potential for debugging the specification by type-checking.
\end{JCS}

\subsection{Structure of the Paper}\label{sec:outline}

We verify the following stack of C program files, listed in dependency order, which link to form an executable.  

\begin{small}\begin{center}
\smallskip\noindent
\begin{tabular}{l@{\hspace*{2ex}}l}\hline
\kw{crypto.h/c}         & library: crypto, malloc, etc (not verified)\\
\iffull\hyperlink{app:RPCdefs.h}{\kw{RPCdefs.h}}\else\kw{RPCdefs.h}\fi          & representation table, event log \\
\iffull\hyperlink{app:hybrids.h}{\kw{RPChybrids.h/c}}\else\kw{hybrids.h/c}\fi     & hybrid wrappers\\
\iffull\hyperlink{app:RPCprot.h}{\kw{RPCprot.h/c}}\else\kw{RPCprot.h/c}\fi        & protocol roles, setup \\
\iffull\hyperlink{app:RPCshim.h}{\kw{RPCshim.h/c}}\else\kw{RPCshim.h/c}\fi        & network attacker interface \\
\kw{RPCattack\_0.c}     & sample attack / application\\\hline
\end{tabular}\smallskip\end{center}
\end{small}

Section~\ref{sec:VCCbackground} introduces features of VCC used in our treatment of symbolic cryptography (the topic of Section~\ref{sec:symbolic}, file \kw{RPCdefs.h}, along with a Coq theory \kw{RPCdefs.v}) and its connection to concrete data
(the topic of Section~\ref{sec:term-bytes}, files \kw{RPChybrids.h} and \kw{RPChybrids.c}).
Section~\ref{sec:semantics} states our assumptions about VCC.
Section~\ref{sec:attack-prog} models symbolic attacks as programs (e.g., \kw{RPCattack\_0.c}) using an API \kw{RPCshim.h}. 
Section~\ref{sec:security-thm} states and proves safety of an example protocol implementation (\kw{RPCprot.h} and \kw{RPCprot.c}).
Section~\ref{sec:additional} summarizes our results including verification of other examples.
Section~\ref{sec:related} covers related work.
Section~\ref{sec:concl} concludes with remarks on limitations and future work.

\begin{SHORT}
A preliminary form of this work was presented at an unrefereed workshop \cite{asa10}.
A technical report~\cite{DGJN11:Guiding-TR} has additional details.
All of the Coq and VCC files are available online.
\FD{Make sure the URL is in the TR. http://fdupress.net/files/confs/CSF11/guiding-source.tar.gz}
\end{SHORT}

\begin{JCS}
This article is based on a conference paper \cite{DGJN11:Guiding},
together with some material extracted from a subsequent paper \cite{FAST11:Csec}.
The updated source files are online at \url{http://fdupress.net/files/journals/guiding/JCS-code.zip}.
\end{JCS}

\section{Background on the VCC Verifier}\label{sec:VCCbackground}

VCC uses an automatic theorem prover to statically check correctness of C code with respect to specifications written as function contracts and other annotation comments. The tool is based on a precise model of multithreaded, shared-memory executions of C programs. In order to verify rich functional specifications without the need for interactive theorem proving and yet scaling to industrial software using idiomatic C, VCC relies on a somewhat intricate methodology for specifications.   This section sketches pertinent features of the model and methodology. 
\begin{FULL}%
Section~\ref{sec:semantics} describes more precisely the way in which our security results rely on VCC.%
\end{FULL}
For details that are glossed over here, see \cite{cohen_vcc:practical_2009,cohen_local_2010} and the tool documentation.
We expect the reader is familiar with C syntax such as macro definitions (\kw{\#define}) and record declarations (\kw{typedef struct}).

VCC's semantics of C is embodied in its verification condition generator (VCG),
\begin{FULL}%
which works by translating annotated C source code to a relatively simple intermediate language (BoogiePL) which itself is equipped with a VCG.  The translation is procedure-modular, that is, each C function body $b$ is verified separately; function calls are interpreted by inlining their specifications.%
\end{FULL}
The VCG reflects a reasoning methodology that includes memory safety and locally checked invariants.  The VCG models preemptive multithreading by interpreting code in terms of its atomic steps, between each of which there may be arbitrary interference on shared state, constrained only by invariants associated with data types declared in the program as explained later.
Atomicity is with respect to sequentially consistent hardware and data types like integers with atomic read and write.
(The methodology has been adapted to reasoning about a weaker memory model, Total Store Order, but that has not yet been implemented in VCC~\cite{CohenSchirmer10}.)
 
Memory blocks are arrays of bytes, but a typed view is imposed in order to simplify reasoning while catering for idiomatic C and standard compilers.  The verifier attempts to associate a type with each pointer dereferenced by the program, and imposes the requirement that distinct pointers reference separate parts of memory.  For example two integers cannot partially overlap. 
 Structs may nest as fields inside other structs, in accord with the declared struct types, but distinct values do not otherwise overlap.
Annotations can specify, however, the re-interpretation of an int as an array of bytes, changing the typestate of a union, and so forth. 

The declaration of a struct type can be annotated with an invariant: a formula that refers to fields of an instance $this$.  
(We often say ``invariant'' for what \correction{is properly called a ``type invariant''}.)
Invariants need not hold of uninitialized objects, 
so there is a boolean ghost field that designates whether the object is \emph{open} or \emph{closed}:
in each reachable state, every closed object should satisfy the invariant associated with its type.
\begin{FULL}%
\emph{Ghost state} is disjoint from the concrete state that exists at runtime; syntactic restrictions ensure that it cannot influence concrete state.  This standard technique appears as ``auxiliary variables'' in \cite{OwickiGries}, as in\ Assumption~\ref{assume:vcc-ghost} in\ Section~\ref{sec:semantics}.%
\end{FULL}

Useful invariants often refer to more than one object, but the point of associating invariants with objects is to facilitate local reasoning: when a field is written, the verifier only needs to check the invariants of relevant objects, owing to \emph{admissibility} conditions that VCC imposes on invariants.  Invariants and other specifications designate an \emph{ownership} hierarchy: if object $o_1$ owns $o_2$ then the invariant of $o_1$ may refer to the state of $o_2$ and thus must be maintained by updates of $o_2$.
The state of a thread is modeled by a ghost object.
An object is \emph{wrapped} if it is owned by the current thread (object) and closed.  
The owner of an object is recorded in a ghost field.
VCC provides notations \kw{unwrap} and \kw{wrap} to open/close an object, with \kw{wrap} also asserting the invariant.

Ownership makes manifest that the invariant for one object $o_1$ may depend on fields of another object $o_2$, so the VCG can check $o_1$'s invariant when $o_2$ is updated. Since hierarchical ownership is inadequate for shared objects like locks, VCC provides another way to make manifest that $o_1$ depends on the state of $o_2$: it allows that $o_1$ maintains a \emph{claim} on $o_2$---a ghost object with no concrete state but an invariant that depends on $o_2$.   Declaring a type to be claimable introduces implicit ghost state used by the VCG to track outstanding claims.
The ghost code to create a claim or store it in a field is part of the annotation provided by the programmer.

The term \emph{invariant} encompasses \emph{two-state} predicates for the before and after states of a state transition.
 In this way, invariants serve as the rely conditions in a form of rely-guarantee reasoning\begin{FULL}\ (compare~\cite{Jones81d}, an early formulation of this concept)\end{FULL}.
Usually two-state invariants are written as ordinary formulas, using the keyword \kw{old} to designate expressions interpreted in the before state.
We say an invariant is \emph{one-state} to mean that it does not depend on the before state.%
\begin{FULL}%
\footnote{To ensure that there is an appropriate interpretation of an invariant in a single state, one of the admissibility conditions is that if an invariant holds for the state transition $\sigma_1$ to $\sigma_2$ it also holds for the stuttering transition $\sigma_2,\sigma_2$ (which thus serves as the one-state interpretation in $\sigma_2$).}%
\end{FULL}

A thread can update an object that it owns, using \kw{unwrap} and \kw{wrap}.  However, in many cases (such as locks), having a single owner is too restrictive, and another mechanism is needed to allow multiple threads to update the object concurrently.  VCC interprets fields marked \kw{volatile} as being susceptible to update by other threads, in accord with the interpretation of the \kw{volatile} keyword by C compilers.  An atomic step is allowed to update a volatile field without opening the object, provided that the object is proved closed and the update maintains the object's two-state invariant (that being the interference condition on which interleaved threads rely).  The standard idiom for locks is that several threads each maintain a claim that the lock is closed, so they may rely on its invariant; outstanding claims prevent even the owner from unwrapping the object. Atomic blocks are explicitly marked as such. An atomic block may make at most one concrete update, to be sound for C semantics, but may update multiple ghost fields.  We do not use \kw{assume} statements in atomic blocks.

\begin{IEEE}
\section{Symbolic Cryptography in VCC}\label{sec:symbolic}

In this section, we introduce the inductive model of symbolic cryptography%
\begin{FULL}, implemented in Coq and shown in full in appendix~\ref{app:RPCdefs.v} for the RPC protocol,\end{FULL}
and show how it is approximated in VCC by first-order logic axioms over uninterpreted function symbols and code manipulating ghost state. 
\subsection{Term Algebra}
We use a standard symbolic model of cryptography, where cryptographic primitives (and further operations such as pairing) are modelled as constructors of an inductive datatype.  Some details are protocol-specific, so for clarity we focus on a simple model adequate for our running example, the RPC protocol. We use ordinary mathematical notation for the following definitions, which are formalized in our Coq development.

\begin{small}
\begin{display}{An algebra of cryptographic terms}
\Category{t_i,k,m,p,a,b}{term}\\
\entry{\textbf{Literal}~{bs}}{with ${bs}$ a byte array}\\
\entry{\textbf{Pair}~t_1~t_2}\\
\entry{\textbf{Hmac}~k~m}
\end{display}
\end{small}

Automated verifiers like VCC support specifications in first-order logic without inductive definitions.  So we use an over-approximation of the term algebra given by uninterpreted function symbols and first-order axioms. The reasoning performed by VCC thus holds for all models of the axioms, in particular for the intended inductive model. The following shows the first-order axioms corresponding to the algebra above.

The VCC \kw{spec()} syntax indicates that all symbols declared within are ghost objects. In particular, it allows the use of the \kw{mathint} type of mathematical integers (in $\mathbb{Z}$).

\begin{SHORT}\pagebreak\end{SHORT}

\begin{vccdisplay}[-2.5]{A first-order model of cryptographic terms}
\lstinputlisting[language=VCC, linerange=DolevYao]{vcc/v7/symcrypt.h}
\end{vccdisplay}

We use the \kw{bytes} type to manipulate whole byte arrays as values, and assume a bijection between finite byte arrays and mathematical integers.  (One such bijection interprets a byte array as an integer, pairs that with its length to account for leading zeroes, and injects the pair into $\mathbb{Z}$.) We will designate the injection from byte arrays to type \kw{bytes} as: \kw{Encode(unsigned char*, unsigned long)}.

The \kw{ispure} keyword is used to specify that a given function is to be interpreted as a total function whose return value depends only on the value of its arguments and memory locations listed in its \kw{reads()} clauses.
Only pure functions can be used in function contracts and assertions, and purity needs to be explicitly specified even for \kw{spec} functions, as they may update ghost state.

We use axioms to state properties of the declared function symbols that cannot easily be expressed using pre and postconditions (injectivity and disjointness, in this case).\footnote{Injectivity could, in general, be expressed as a postcondition, but VCC imposes syntactic restrictions on the postconditions of pure functions to ensure that they are total and computable.} 
Those axioms are separately proved in Coq to hold about the intended, inductive model of cryptography.
The \kw{theorem} notation is a simple macro that generates a VCC \kw{axiom}; its first argument, ignored by VCC, is the name of the corresponding Coq theorem.

\subsection{Events and Log}
Much as in prior work \cite{bhargavan_modular_2010}, we use a global log of events to express the wanted correspondence properties. Events themselves are defined using a protocol-specific algebra (see below).

The hashkey usage and some top-level events are specific to the RPC protocol, but some constructors are of general use and are needed for most protocols.  The top-level of events always includes a \textbf{New} event, logging the intended usage of freshly generated bytestrings. In particular, the \textbf{AttackerGuess} models that the corresponding term represents a bytestring known to the attacker, because it was provided as a starting parameter, or because it represents a fixed bytestring appearing in the protocol specification (e.g. a tag, or a fixed format string).

\begin{small}
\begin{display}{Algebra of events for RPC}
\Category{hu}{hashkey usage}\\
\entry{\textbf{KeyAB}~a~b}\\
\Category{us}{usage}\\
\entry{\textbf{AttackerGuess}}\\
\entry{\textbf{HashKey}~hu}\\
\Category{ev}{event}\\
\entry{\textbf{New}~t~us 
  \mid \textbf{Bad}~a
  \mid \textbf{Request}~a~b~t
  \mid \textbf{Response}~a~b~t_1~t_2}{}
\end{display}
\end{small}

As with the cryptographic terms, a first-order approximation of this algebra is given to VCC using uninterpreted functions and axioms. The log itself, intended as the set of events that have occurred so far, is defined next as a structure containing one set for each kind of event. We use boolean maps to model sets.  The VCC notation is similar to that of arrays, e.g., \kw{bool B[mathint]} declares \kw{B} to be a boolean-valued total function on the integers.

\begin{FULL}%
Logical formulas have type \kw{bool}, e.g., the expression \kw{Request[a][b][t]} can be used as an assertion saying that this event has occurred. 
We use two-state invariants to express that the log can only grow (see the ``Stability'' group of invariants in the code displayed below).  We use a one-state invariant to express that bytestrings, and in particular keys, should only be given one usage. Other protocol-specific one-state invariants could be added. We call them ``Miscellaneous conditions'', and will say that a log is \emph{good}, or \emph{valid}, when its miscellaneous conditions hold.%
\end{FULL}

\begin{vccdisplay}{Encoding of the log in VCC (\kw{RPCdefs.h})}
\lstinputlisting[language=VCC,linerange=Log]{vcc/RPC/RPCdefs.h}
\end{vccdisplay}

\begin{SHORT}
Logical formulas have type \kw{bool}, e.g., the expression \kw{Request[a][b][t]} can be used as an assertion saying that this event has occurred.
We use two-state invariants to express that the log can only grow (see the ``Stability'' group of invariants in the code displayed above).  We use a one-state invariant to express that bytestrings, and in particular keys, should only be given one usage. Other protocol-specific one-state invariants could be added. We call them ``Miscellaneous conditions'', and will say that a log is \emph{good}, or \emph{valid}, when its miscellaneous conditions hold.  
\end{SHORT}
We also introduce macros \kw{valid_log} and \kw{stable_log}, which expand to both invariant blocks, to simplify the expression of certain properties; \kw{valid_log} is used in the inversion principle for the \textit{Pub()} predicate.

\subsection{Inductive Predicates for Cryptography}\label{sec:inductive}

We use inductive predicates to express the correct usage of cryptographic primitives, as specified by a given protocol. In particular, we define a predicate \textit{Pub()} that holds on all terms that can be published without compromising the protocol's goals. We also define a \textit{Bytes()} predicate that holds on byte arrays an honest protocol participant is allowed to build. We ensure by definition that \textit{Bytes()} holds for all terms on which \textit{Pub()} holds. Both of these predicates, and all intermediate predicates used in their definition, are actually functions of the log. We write $\mathcal{L}\vdash P$ to say $P$ holds in log $\mathcal{L}$. The following shows an excerpt of the inductive rules defining the \textit{Pub()} predicate.
\begin{FULL}%
Those rules are very similar to those seen in previous work~\cite{bhargavan_modular_2010}, in which they are described in more detail. The main difference is that we make the log explicit in the definitions, avoiding the need for customised definitions of safety.%
\end{FULL}

\begin{small}
\begin{display}{Some inductive predicates for cryptography}
\staterule{MACSays KeyAB Request}
  {\textbf{New}~k~(\textbf{HashKey}~(\textbf{KeyAB}~a~b))~\in~\mathcal{L}\\
    \textbf{Request}~req~a~b~\in~\mathcal{L}\\
    m~=~\textbf{Pair}~(\textbf{Literal}~TagRequest)~req}
  {\mathcal{L}~\vdash~\textit{MACSays}~k~m}
\\[3ex]
\staterule{MACSays KeyAB Response}
  {\textbf{New}~k~(\textbf{HashKey}~(\textbf{KeyAB}~a~b))~\in~\mathcal{L}\\
    \textbf{Response}~req~resp~a~b~\in~\mathcal{L}\\
    \quad m~=~\textbf{Pair}~(\textbf{Literal}~TagResponse)~(\textbf{Pair}~req~resp)}
  {\mathcal{L}~\vdash~\textit{MACSays}~k~m}
\\[3ex]
\staterule{Pub AttackerGuess}
  {\textbf{New}~(\textbf{Literal}~b)~\textbf{AttackerGuess}~\in~\mathcal{L}}
  {\mathcal{L}~\vdash~\textit{Pub}~(\textbf{Literal}~b)}
\\[3ex]
\iffull
\staterule{Pub HmacKey KeyAB}
  {\textbf{New}~(\textbf{Literal}~bk)~(\textbf{HashKey}~(\textbf{KeyAB}~a~b))~\in~\mathcal{L}\\
    \textbf{Bad}~a~\in~\mathcal{L}~\vee~\textbf{Bad}~b~\in~\mathcal{L}}
  {\mathcal{L}~\vdash~\textit{Pub}~(\textbf{Literal}~bk)}
\\[3ex]
\staterule{Pub Pair}
  {\mathcal{L}~\vdash~\textit{Pub}~t_1\quad
    \mathcal{L}~\vdash~\textit{Pub}~t_2}
  {\mathcal{L}~\vdash~\textit{Pub}~(\textbf{Pair}~t_1~t_2)}
\\[3ex]
\fi
\staterule{Pub Hmac}
  {\mathcal{L}~\vdash~\textit{MACSays}~k~m\\
    \mathcal{L}~\vdash~\textit{Bytes}~k\\
    \mathcal{L}~\vdash~\textit{Pub}~m}
  {\mathcal{L}~\vdash~\textit{Pub}~(\textbf{Hmac}~k~m)}
\quad
\staterule{Pub Hmac Pub}
  {\mathcal{L}~\vdash~\textit{Pub}~k\quad
    \mathcal{L}~\vdash~\textit{Pub}~m}
  {\mathcal{L}~\vdash~\textit{Pub}~(\textbf{Hmac}~k~m)}
\iffull
\\[3ex]
\staterule{Bytes Literal}
  {}
  {\mathcal{L}~\vdash~\textit{Bytes}~(\textbf{Literal}~b)}
\quad
\staterule{Bytes Pair}
  {\mathcal{L}~\vdash~\textit{Bytes}~t_1\\
    \mathcal{L}~\vdash~\textit{Bytes}~t_2}
  {\mathcal{L}~\vdash~\textit{Bytes}~(\textbf{Pair}~t_1~t_2)}
\\[3ex]
\staterule{Bytes Hmac}
  {\mathcal{L}~\vdash~\textit{MACSays}~k~m\\
    \mathcal{L}~\vdash~\textit{Bytes}~k\\
    \mathcal{L}~\vdash~\textit{Bytes}~m}
  {\mathcal{L}~\vdash~\textit{Bytes}~(\textbf{Hmac}~k~m)}
\quad
\staterule{Bytes Hmac Pub}
  {\mathcal{L}~\vdash~\textit{Pub}~k\quad
    \mathcal{L}~\vdash~\textit{Pub}~m}
  {\mathcal{L}~\vdash~\textit{Bytes}~(\textbf{Hmac}~k~m)}
\fi
\end{display}
\end{small}

In order to simplify the notations in VCC, we do not make the log an explicit argument in the predicates' VCC declaration.
Instead, we express that the functions intended to model the predicates depend on the state, by marking them with the \kw{reads(set_universe())} contract, stating that the value returned by the function may depend on the set of pointers that contains all addresses in the state.
This makes the program state an implicit parameter to the function; later we use axioms to frame the function's dependency on the state more precisely.
We can then use axioms to give the intended meaning of these symbols.
In particular, we prove in Coq that our predicate symbols are monotonic functions of the log and import this fact as an axiom in VCC (see the theorem \kw{Pub Monotonic} below).
To that end we give the following definitions.
The notation $\kw{in_state}(S,e)$ denotes the value of expression $e$ in the state $S$.


\begin{vccdisplay}{Log stability between explicit states}
\lstinputlisting[language=VCC,linerange=LogStateStability]{vcc/RPC/RPCdefs.h}
\end{vccdisplay}

Here are the VCC axioms for the \kw{Pub()} predicate and some example theorems.

\begin{vccdisplay}{VCC axiomatic definition for \kw{Pub()} (excerpt)}
\lstinputlisting[language=VCC,linerange=PubDefs]{vcc/RPC/RPCdefs.h}
\end{vccdisplay}
\begin{vccdisplay}{Inversion theorems (excerpt)}
\lstinputlisting[language=VCC,linerange=AdditionalRPCDefs]{vcc/RPC/RPCdefs.h}
\end{vccdisplay}

The axioms introduced using \kw{rule} correspond to the inductive definition rules shown earlier.  Again, the axioms introduced using \kw{theorem} correspond to results that are proved to hold in the model inductively defined in Coq.
Similar axioms are used to define \kw{Bytes()} and \kw{MACSays()}, as well as theorems similar to those used in F7~\cite{bhargavan_modular_2010}.

\begin{FULL}%
We also separately prove in Coq (and import as axioms in VCC) some weak secrecy theorems, expressing under which conditions a key or nonce can become known to the attacker.
In the case of RPC, the weak secrecy theorem simply states that a key shared between principals $a$ and $b$ can only be public if either $a$ or $b$ is compromised, as follows.%
\[\begin{prog}
(\textbf{New}~k~(\textbf{KeyAB}~a~b) \in \log) \wedge (\log \vdash \textbf{Pub}(k)) \Rightarrow {} \\ \quad (\textbf{Bad}~a \in \log) \vee (\textbf{Bad}~b \in \log)
\end{prog}\]

The verification of correspondence assertions in VCC may make use of this theorem, which holds in any system that enforces the log invariants---and in particular of any system verified using VCC, to establish the correspondence properties only in terms of logged events.

\end{FULL}
\end{IEEE}

\begin{JCS}
\section{Symbolic Cryptography in VCC}\label{sec:symbolic}

The original work on cryptographic invariants in F7 \cite{bhargavan_modular_2010} introduces inductive definitions simply by listing Horn clauses.
In our work with VCC, we express the symbolic algebra and its cryptographic invariants as explicit Coq definitions.
We adapt the presentation of those Coq definitions from a previous paper~\cite{FAST11:Csec},
which introduced a more systematic representation of cryptographic invariants,
to the HMAC-based RPC example studied in this article.

This section describes a type \coqkw{term} of symbolic cryptographic expressions,
a type \coqkw{log} of sets of events during runs of a protocol,
and a type \coqkw{level}, either \coqkw{Low} or \coqkw{High}.
Given these types, we make an inductive definition of a predicate \coqkw{Level l t L},
meaning that the term \coqkw{t} may arise at level \coqkw{l} after the events in log \coqkw{L} have happened.
The set of terms at level \coqkw{Low} is an upper bound on any attacker's knowledge,
while the set of terms at level \coqkw{High} is an upper bound on any principal's knowledge.
(The set of \coqkw{High} terms is a strict superset of the \coqkw{Low} terms.)
We make these definitions in the Coq proof assistant, and use it to check security theorems.
Subsequently, we import the definitions and theorems into VCC, confident in their soundness.

We present the Coq definitions for our authenticated RPC example,
along with their encoding in VCC, which makes use of the newly implemented support for inductive datatypes and functions.
The use of datatypes lets VCC generate first-order axioms that are much more efficient than those we wrote by hand in prior work~\cite{DGJN11:Guiding},
and provides better type-checking to help debug complicated protocol specifications.

\subsection{Terms and Usages}
First, we define the \coqkw{term} type,
with constructors to build literal terms from bytestrings,
to injectively pair two terms (the $\cdot | \cdot$ operation),
to compute keyed hashes,
and, for generality, to perform symmetric authenticated encryption.
(To accommodate other protocols, we may extend the type with constructors for other standard cryptographic primitives, such as asymmetric encryption and signature.)

We define an auxiliary type \coqkw{usage}, whose values describe the purposes of freshly generated bytestrings of the protocol.
These may be guesses generated by the attacker,
or protocol keys, or nonces sent as messages to help us specify secrecy properties.
In this protocol, we only have one kind of keys for keyed hashes,
and all other usage types (\coqkw{nonceUsage} and \coqkw{sencKeyUsage}) are empty.
\begin{FULL}%
Appendix~\ref{app:OR} shows the narration and usage definitions for a variant of the Otway-Rees key exchange protocol we also verify,
showing an example involving several distinct key usages.
The inductive definitions shown in~\cite{FAST11:Csec} for an encryption-based variant of RPC also show nonce usages, as well as multiple symmetric encryption key usages.%
\end{FULL}
\begin{vccdisplay}{Coq and VCC definitions for terms and usages}
\vspace{-1em}
\begin{multicols}{2}
\begin{lstlisting}[language=coq]
Inductive term :=
  | Literal: (bs: bytes)
  | Pair: (t1 t2: term)
  | Hmac: (k m: term)
  | SEnc: (k p: term).

Inductive hmacKeyUsage :=
  | U_KeyAB: (a b: term).

Inductive usage :=
  | AttackerGuess
  | Nonce: nonceUsage
  | HmacKey: hmacKeyUsage
  | SEncKey: sencKeyUsage.
\end{lstlisting}
\columnbreak
\begin{lstlisting}[language=vcc]
_(datatype term {
    case Literal(bytes)
    case Pair(term,term)
    case Hmac(term,term)
    case SEnc(term,term) })

_(datatype hmacKeyUsage {
    case U_KeyAB(term a, term b) })

_(datatype usage {
    case AttackerGuess()
    case Nonce(nonceUsage u)
    case HmacKey(hmacKeyUsage u)
    case SEncKey(sencKeyUsage u) })
\end{lstlisting}
\end{multicols}
\vspace{-.8em}
\end{vccdisplay}

\subsection{Events and Log}
Next, we introduce the \coqkw{log} type as being a set of events, where there are four constructors of the \coqkw{event} type:
(1) an event \coqkw{New (Literal bs)} \coqkw{u} means that the fresh bytestring \coqkw{bs} has one of the usages \coqkw{u}; 
(2) an event \coqkw{Request a b req} means that client \coqkw{a} intends to send server \coqkw{b} the request \coqkw{req};
(3) an event \coqkw{Response a b req resp} means that server \coqkw{b} has accepted the request \coqkw{req} from client \coqkw{a} and intends to reply with response \coqkw{resp};
(4) an event \coqkw{Bad p} means that any key known to principal \coqkw{p} is compromised.
We also define a predicate \coqkw{Good L} to mean that the \coqkw{New} events in log \coqkw{L} ascribe a unique usage to each nonce or key, and apply only to bytestring literals.

In VCC, we model the type of sets using boolean maps, and set membership is simple map application, denoted using square brackets.
Functions preceded by the \kw{def} keyword must be pure terminating functions and their body is available as an expression when verifying other functions, allowing the omission of precise contracts, at the cost of a slight decrease in performance.

\noindent\begin{minipage}{\textwidth}
\begin{vccdisplay}{Coq and VCC definitions for events and logs}
\vspace{-1em}
\begin{multicols}{2}
\begin{lstlisting}[language=coq]
Inductive ev :=
  | New: (t: term) (u: usage)
  | Request: (a b req: term)
  | Response: (a b req resp: term)
  | Bad: (p: term).

Definition log := set event.
Definition Logged (e: ev) (L: log) :=
  set_In e L.

Definition log_leq (L L': log) :=
  forall x, Logged x L -> Logged x L'.

Definition Good (L: log) :=
  (forall t u, Logged (New t u) L ->
    exists bs, t = Literal bs) /\
  (forall t u1 u2,
    Logged (New t u1) L ->
    Logged (New t u2) L ->
    u1 = u2).
\end{lstlisting}
\columnbreak
\begin{lstlisting}[language=vcc]
_(datatype event {
    case New(term,usage)
    case Bad(term)
    case Request(term,term,term)
    case Response(term,term,term,term) })

_(typedef \bool log[event])

_(def \bool leq_log(log L1,log L2)
  { return \forall event e;
      L1[e] ==> L2[e]; })

_(def \bool valid_log(log L)
  { return
      (\forall term t; usage u; L[New(t,u)] ==>
         \exists bytes b; t == Literal(b)) &&
      (\forall term t; usage u1,u2;
         L[New(t,u1)] ==>
         L[New(t,u2)] ==>
         u1 == u2); })
\end{lstlisting}
\end{multicols}
\vspace{-.5em}
\end{vccdisplay}
\end{minipage}

A central idea of cryptographic invariants is that each key usage has an associated \emph{payload property},
which relates keys and payloads to which honest principals can apply the corresponding cryptographic primitive.
The payload property \coqkw{RPCKeyABPayload a b m L} says that
a key shared between \coqkw{a} and \coqkw{b}%
may be used
to compute the keyed hash of a payload \coqkw{m}
when \coqkw{m} is a pair composed of a constant tag \coqkw{tagRequest},
and a request from \coqkw{a} to \coqkw{b}
on which the \coqkw{Request} event has been logged in \coqkw{L},
or when it is a pair composed of a constant tag \coqkw{tagResponse},
together with the injective pairing of a request \coqkw{req}%
and a response \coqkw{resp}
on which the \coqkw{Response} event has been logged in \coqkw{L}.
The formatting conditions are abstracted in the \coqkw{Requested} and \coqkw{Responded} predicates.
We combine payload properties in the definition of general preconditions,
such as the \coqkw{canHmac} predicate below, which serves as a precondition,
in code, to the keyed hashing function when called by honest participants.
\begin{vccdisplay}{Coq and VCC definitions for payload conditions}
\vspace{-1em}
\begin{multicols}{2}
\begin{lstlisting}[language=coq]
Definition KeyAB a b k (L: log) :=
  Logged (New k (HmacKey (U_KeyAB a b))) L.

Definition KeyABPayload a b m (L: log) :=
  (exists req,
    Requested m req /\
    Logged (Request a b req) L) \/
  (exists req, exists resp,
    Responded m req resp /\
    Logged (Response a b req resp) L).

Definition canHmac (k m: term) (L: log) :=
  (exists a, exists b,
     KeyAB a b k L /\
     KeyABPayload a b p L).
\end{lstlisting}
\columnbreak
\begin{lstlisting}[language=vcc]
_(def \bool KeyAB(term a,term b,term k,log L)
  { return
      L[New(k,HmacKey(U_KeyAB(a,b)))]; })

_(def \bool KeyABPayload(term a,term b,
                        term m,log L)
  { return
      (\exists term req; Requested(m,req) &&
        L[Request(a,b,req)]) ||
      (\exists term req,resp;
        Responded(m,req,resp) &&
        L[Response(a,b,req,resp)]); })

_(def \bool canHmac(term k,term m,log L)
  { return
      \exists term a,b;
         KeyAB(a,b,k,L) &&
         KeyABPayload(a,b,m,L); })
\end{lstlisting}
\end{multicols}
\vspace{-.8em}
\end{vccdisplay}

Another central idea is that each nonce or key has a \emph{compromise condition},
which needs to be fulfilled before a literal given that usage can be released to the attacker.
Implicitly, bytestrings with usage \coqkw{AttackerGuess} are always known to the attacker.
Our next predicate defines the compromise condition for HMAC keys.
In the general case where there are several usages, the \coqkw{hmacComp} predicate
gathers them all in a disjunction that serves as an abstract way of expressing key compromise.

\begin{vccdisplay}{Coq and VCC definitions for compromise conditions}
\vspace{-1em}
\begin{multicols}{2}
\begin{lstlisting}[language=coq]
Definition KeyABComp a b k (L: log) :=
  Logged (Bad a) L \/
  Logged (Bad b) L.

Definition hmacComp (k: term) (L: log) :=
  (exists a, exists b,
    KeyAB a b k L /\
    KeyABComp a b k L).
\end{lstlisting}
\columnbreak
\begin{lstlisting}[language=vcc]
_(def \bool KeyABComp(term a,term b,log L)
  { return L[Bad(a)] || L[Bad(b)]; })

_(def \bool hmacComp(term k,log L)
  { return
      \exists term a,b;
        KeyAB(a,b,k,L) &&
        KeyABComp(a,b,L); })
\end{lstlisting}
\end{multicols}
\vspace{-.8em}
\end{vccdisplay}

\subsection{Inductive \coqkw{Level} Predicate}\label{sec:inductive}
Given these auxiliary predicates, we now define the \coqkw{Level} predicate.
We intend that given a log \coqkw{L},
any term \coqkw{t} sent or received on the network satisfies \coqkw{Level Low t L},
while if \coqkw{t} is data manipulated internally by principals, we must have \coqkw{Level High t L}.
(The \coqkw{Level} predicate consolidates both the \coqkw{Pub} and \coqkw{Bytes} predicates from \cite{DGJN11:Guiding}; specifically, \coqkw{Level Low} is a predicate equivalent to \coqkw{Pub}, and \coqkw{Level High} is a predicate equivalent to \coqkw{Bytes}.)
It easily follows from the definition that any term satisfying \coqkw{Level Low} also satisfies \coqkw{Level High}
(but not the converse,
because for example uncompromised keys and nonces satisfy  \coqkw{Level High} but not  \coqkw{Level Low} until they are compromised).
We also prove that \coqkw{Level l} is a monotonic function of its log argument for all \coqkw{l}, which will help greatly when using the definitions in VCC.
Additionally, we also prove slight variants of some of the rules that will be used to help VCC and Z3 efficiently instantiate quantified variables.
\begin{vccdisplay}{Coq inductive definitions for the \coqkw{Level} predicate}
\begin{lstlisting}[language=coq]
Inductive level := Low | High.
Inductive Level: level -> term -> log -> Prop :=
  | Level_AttackerGuess: forall l bs L, (* AttackerGuesses are always Low *)
    Logged (New (Literal bs) AttackerGuess) L ->
    Level l (Literal bs) L

  | Level_Nonce: forall l bs L nu, (* Nonces are Low when compromised *)
    Logged (New (Literal bs) (Nonce nu)) L ->
    (l = Low -> nonceComp (Literal bs) L) ->
    Level l (Literal bs) L

  | Level_HMacKey: forall l bs L hu, (* HMacKeys are Low when compromised *)
    Logged (New (Literal bs) (HMacKey hu)) L ->
    (l = Low -> hmacComp (Literal bs) L) ->
    Level l (Literal bs) L

  | Level_SEncKey: forall l bs L su, (* SEncKeys are Low when compromised *)
    Logged (New (Literal bs) (SEncKey su)) L ->
    (l = Low -> sencComp (Literal bs) L) ->
    Level l (Literal bs) L

  | Level_Pair: forall l t1 t2 L, (* Pairs have same level as their components *)
    Level l t1 L ->
    Level l t2 L ->
    Level l (Pair t1 t2) L

  | Level_Hmac: forall l k m L, (* HMAC with payload matching payload property *)
    canHmac k m L ->
    Level l m L ->
    Level l (Hmac k m) L

  | Level_Hmac_Low: forall l k m L, (* HMAC with compromised or Low key *)
    Level Low k L ->
    Level Low m L ->
    Level l (Hmac k m) L

  | Level_SEnc: forall l l' k p L, (* SEnc with plaintext matching payload property *)
    canSEnc k p L ->
    Level l' p L ->
    Level l (SEnc k p) L

  | Level_SEnc_Low: forall l k p L, (* SEnc with compromised or Low key *)
    Level Low k L ->
    Level Low p L ->
    Level l (SEnc k p) L

Theorem Low_High: forall t L, Level Low t L -> Level High t L.
Theorem Level_Positive: forall l t L L', log_leq L L' -> Level l t L -> Level l t L'.
\end{lstlisting}
\end{vccdisplay}

Unlike all previously presented definitions, this predicate and all the theorems to come are not imported as structural definitions into VCC.
Instead, the Level predicate is declared as an uninterpreted function;
and each rule and theorem is separately imported as a first-order axiom.
They are imported using the \kw{rule} or \kw{theorem} macros,
that both expand to the \kw{axiom} keyword, erasing the name.

The following snippet of VCC code shows the VCC axiomatisation corresponding to the subset of the rules above used in the RPC protocol.
The terms in curly braces are triggers, that guide the SMT-solver when instantiating the quantified variables, and can significantly improve performance (and sometimes induce termination), at the cost of some loss in precision.
The triggers shown here should be good enough for most protocols, but the reader willing to experiment should read~\cite{Moskal:ProgrammingWithTriggers}.

\begin{vccdisplay}{Axiomatic VCC definition for \coqkw{Level} (excerpt) and related theorems}
\begin{lstlisting}[language=vcc]
_(datatype level {
    case Low()
    case High() })
_(abstract \bool Level(level l,term t,log L))

_(rule(Level_AttackerGuess)
  \forall level l; bytes bs; log L;
    L[New(Literal(bs),AttackerGuess())] ==>
    Level(l,Literal(bs),L))

_(rule(Level_HmacKey)
  \forall level l; bytes bs; log L; hmacKeyUsage hu;
    { HmacKey(hu), Level(l,Literal(bs),L) }
    L[New(Literal(bs),HmacKey(hu))] ==>
    (l == Low() ==> hmacComp(Literal(bs),L)) ==>
    Level(l,Literal(bs),L))

_(rule(Level_Pair)
  \forall level l; term t1,t2; log L;
    { Level(l,Pair(t1,t2),L) }
    Level(l,t1,L) ==>
    Level(l,t2,L) ==>
    Level(l,Pair(t1,t2),L))

_(rule(Level_Hmac)
  \forall level l; term k,m; log L;
    canHmac(k,m,L) ==>
    Level(l,m,L) ==>
    Level(l,Hmac(k,m),L))

_(rule(Level_Hmac_Low)
  \forall level l; term k,m; log L;
    Level(Low(),k,L) ==>
    Level(Low(),m,L) ==>
    Level(l,Hmac(k,m),L))

_(theorem(Low_High)
   \forall term t; log L;
      Level(Low(),t,L) ==>
      Level(High(),t,L))

_(theorem(Level_Positive)
  \forall log L1,L2; level l; term t;
    leq_log(L1,L2) ==>
    Level(l,t,L1) ==>
    Level(l,t,L2))

_(theorem(Pair_Level)
   \forall level l; term t1,t2; log L;
     Level(l,Pair(t1,t2),L) ==>
     Level(l,t1,L) && Level(l,t2,L))
\end{lstlisting}
\end{vccdisplay}

We state secrecy properties of the protocol as consequences of the invariants respected by the code.
We prove in the following theorem that HMAC keys are kept secret unless their compromise condition is fulfilled.
We actually state the contrapositive: that if \coqkw{Level Low} holds on the key (intuitively, if the key is not secret),
then one of the principals that shares it is \coqkw{Bad}.
The proof is an almost direct application of the inversion principle for the \coqkw{Level_HmacKey} inductive rule above: the only way for a
literal with an HMAC key usage to be \coqkw{Low} is for its compromise condition to hold.

\noindent\begin{minipage}{\textwidth}
\begin{vccdisplay}{Secrecy theorems for HMAC keys in Coq and VCC}
\vspace{-1em}
\begin{multicols}{2}
\begin{lstlisting}[language=coq]
Theorem WeakSecrecyKeyAB: forall a b k L,
  Good L ->
  KeyAB a b k L ->
  Level Low k L ->
  Logged (Bad a) L \/ Logged (Bad b) L.
\end{lstlisting}
\columnbreak
\begin{lstlisting}[language=vcc]
_(theorem(WeakSecrecyKeyAB)
  \forall term k,a,b; log L;
    valid_log(L) ==>
    KeyAB(a,b,k,L) ==>
    Level(Low(),k,L) ==>
    L[Bad(a)] || L[Bad(b)])
\end{lstlisting}
\end{multicols}
\vspace{-.8em}
\end{vccdisplay}
\end{minipage}\smallskip

This secrecy property states the absence of a direct flow of a key to the opponent, unless one of the associated principals is compromised.
We do not address here how to show noninterference properties, i.e.\ the absence of indirect flows.

Finally, we state a simple inversion theorem for HMACs (and similarly for all cryptographic primitives if necessary), stating that, in a good log, an HMAC computed using a valid key (which can be compromised, but has to have a correct usage) can only be \coqkw{High} if either \coqkw{canHmac} holds on the key and payload, or the key is compromised.

\begin{vccdisplay}{Coq and VCC inversion theorem for HMACs}
\vspace{-1em}
\begin{multicols}{2}
\begin{lstlisting}[language=coq]
Theorem Hmac_Inversion: forall hu m k L,
  Good L ->
  Logged (New k (HmacKey hu)) L ->
  Level High (Hmac k m) L ->
  canHmac k m L \/ hmacComp k L.
\end{lstlisting}
\columnbreak
\begin{lstlisting}[language=vcc]
_(theorem(Hmac_Inversion)
  \forall hmacKeyUsage hu; term m,k; log L;
    valid_log(L) ==>
    L[New(k,HmacKey(hu))] ==>
    Level(High(),Hmac(k,m),L) ==>
    canHmac(k,m,L) || hmacComp(k,L))
\end{lstlisting}
\end{multicols}
\vspace{-1em}
\end{vccdisplay}

We embed the assertions from the narration at the start of this section within the code
at the points that the request and response messages have been validated;
to verify these assertions we simply rely on the postcondition of MAC verification, and the definition of \coqkw{canHmac} and \coqkw{hmacComp}.

\subsection{Correspondence Theorems}

However, it may be desirable to prove directly on the abstract model that the defined usages, along with their payload and compromise conditions do indeed guarantee desired authentication properties, regardless of the implementation, both in terms of protocol narration and of concrete implementation (for example, to prove security properties of the model before implementing the protocol).

To do so, we prove the following two correspondence theorems in Coq.
The first states that if there is a public HMAC keyed by a key with usage \coqkw{KeyAB},
and starting with tag \coqkw{tagRequest},
then either \coqkw{Request} has been logged on the rest of the payload,
or one of the principals involved in the exchange is compromised.
The second states a similar result for responses.

\noindent\begin{minipage}{\textwidth}
\begin{vccdisplay}{Coq theorems for correspondence properties}
\vspace{-1em}
\begin{multicols}{2}
\begin{lstlisting}[language=coq]
Theorem AuthenticationRequest: forall a b req k L,
  Good L ->
  KeyAB a b k L ->
  Level Low (Hmac k (Pair tagRequest req)) L ->
  Logged (Request a b req) L \/
  Logged (Bad a) L \/
  Logged (Bad b) L.
\end{lstlisting}
\columnbreak
\begin{lstlisting}[language=coq]
Theorem AuthenticationResponse: forall a b req resp k L,
  Good L ->
  KeyAB a b k L ->
  Level Low (Hmac k (Pair tagResponse (Pair req resp))) L ->
  Logged (Response a b req resp) L) \/
  Logged (Bad a) L \/
  Logged (Bad b) L.
\end{lstlisting}
\end{multicols}
\vspace{-.5em}
\end{vccdisplay}
\end{minipage}
\end{JCS}

\section{Representation Table and Hybrid Wrappers}\label{sec:term-bytes}
\subsection{The Representation Table}
Symbolic models of cryptography generally assume that two distinct symbolic terms yield two distinct byte strings, and that fresh literals cannot be guessed by an attacker.
The intent is to use such a model with cryptographic operations that, in the computational model, have a negligible probability of collision.
Verification in the symbolic model is a way of ruling out a well-defined class of attacks,
which do not depend on collisions or lucky guesses.

Prior work on cryptographic software in {F\#},
for example \cite{BFGT08:VerifiedInteroperableImplementationsOfSecurityProtocols,DBLP:journals/toplas/BengtsonBFGM11},
relies on type abstraction to verify protocol code, by reasoning in terms of purely symbolic libraries,
which satisfy these assumptions,
instead of concrete libraries, which do not.
In the absence of type abstraction in C,
we must verify protocol code linked with concrete cryptographic algorithms on byte strings.
Our aim remains to verify against an attacker in the symbolic model.
To do so, we instrument the program with specification code that maintains a \emph{representation table},
which tracks the correspondence between concrete byte strings and symbolic terms.
We intercept all calls to cryptographic functions with ghost code to update the representation table.
We say a \emph{collision} occurs when the table associates a single byte string with two distinct symbolic terms.

\vspace{0.2cm}\hspace{-0.35cm}\begin{minipage}{4.5cm}
For example, suppose $x$ and $y$ are two distinct bytestrings that have the same HMAC, $h$, under a key $k$.
After the first call to \kw{hmac()} the table looks like this:
\end{minipage}
\hspace{0.4cm}\vspace{-0.5cm}
\begin{minipage}{3cm}
\begin{footnotesize}
\begin{center}
\begin{tabular}{|l|l|}
\hline
Bytestring	& Term\\\hline
$k$		& \textbf{Literal}~$k$\\\hline
$x$		& \textbf{Literal}~$x$\\\hline
$y$		& \textbf{Literal}~$y$\\\hline
$h$		& \textbf{Hmac}~$k~x$\\\hline
\end{tabular}
\end{center}
\end{footnotesize}
\end{minipage}
\vspace{0.7cm}

When computing the second HMAC, our instrumented \kw{hmac()} function tries to insert the freshly computed $h$ and the corresponding term \textbf{Hmac}~$k~y$ in the table, but detects that $h$ is already associated with a distinct term \textbf{Hmac}~$k~x$.

We make the absence of such collisions an explicit hypothesis in our specification by assuming,
via an \kw{assume} statement in the ghost code updating the table, that a collision has not occurred.  This removes from consideration any computation following a collision, as is made precise in 
Section~\ref{sec:semantics}.  
%
%
We treat the event of the attacker guessing a non-public value in a similar way;
we assume it does not happen, using an \kw{assume} statement.
In this way, we prove symbolic security properties of the C code.  
A separate argument may be made that such collisions only happen with low probability.

%
%

%
%
%
%
%
%
%

\begin{IEEE}
Like the log, the representation table, given next, is a structure containing maps.
\begin{FULL}
Its VCC definition is shown below.
\pagebreak
\end{FULL}

\begin{vccdisplay}[-2.5]{The representation table in VCC}
\lstinputlisting[language=VCC,linerange=Table]{vcc/v7/table.h}
\end{vccdisplay}

We use two maps to store the bijection between \kw{bytes}, which are byte string
values (not pointers), and \kw{terms}
(e.g., the byte string \kw{b} corresponds to the algebraic term \kw{B2T[b]}).
Two-state invariants express that the table can only grow.
There is an ownership invariant: the representation table always owns the log. This means that whenever \kw{&rep} is closed, \kw{&log} has to be closed itself, so the invariants for the representation table can depend on values in the log in accord with the VCC ownership methodology.
\begin{FULL}
This ownership relation also enables us to make only the table object claimable (using the \kw{vcc(claimable)} attribute on the structure declaration), releasing some of the pressure on the prover in later stages. A thread holding a claim on the table will be able to show, through the ownership relation, that the log is also closed.
\end{FULL}
\end{IEEE}

\begin{JCS}
Like the log, we define the table as a ghost object, this time a functional record.

\begin{vccdisplay}{Representation tables in VCC}
\begin{lstlisting}[language=vcc]
_(record table {
    \bool DefinedB[bytes];
    term B2T[bytes];

    \bool DefinedT[term];
    bytes T2B[term]; })

_(def \bool valid_table(log L,table T)
  { return
      (\forall bytes b;
        T.DefinedT[Literal(b)] ==>
        T.T2B[Literal(b)] == b) &&
      (\forall term t;
        T.DefinedT[t] ==>
        Level(High(),t,L)) &&
      (\forall bytes b;
        T.DefinedB[b] ==>
        T.T2B[T.B2T[b]] == b) &&
      (\forall term t;
        T.DefinedT[t] ==>
        T.B2T[T.T2B[t]] == t) &&
      (\forall bytes b;
        T.DefinedB[b] ==>
        T.DefinedT[T.B2T[b]]) &&
      (\forall term t;
        T.DefinedT[t] ==>
        T.DefinedB[T.T2B[t]]); })

_(def \bool leq_table(table T1,table T2)
  { return
      (\forall bytes b;
        T1.DefinedB[b] ==>
        T2.DefinedB[b]) &&
      (\forall bytes b;
        T1.DefinedB[b] ==>
        T1.B2T[b] == T2.B2T[b]) &&
      (\forall term t;
        T1.DefinedT[t] ==>
        T2.DefinedT[t]) &&
      (\forall term t;
        T1.DefinedT[t] ==>
        T1.T2B[t] == T2.T2B[t]); })
\end{lstlisting}
\end{vccdisplay}

We use two maps to store the bijection between \kw{bytes}, which are bytestring values (not pointers), and \kw{terms}.
The predicate \kw{valid_table} expresses that the \kw{B2T} and \kw{T2B} maps of a table do indeed represent a bijection, and also adds conditions relating elements stored in the table to a particular log.
We also define an order on tables, which is the standard order on pairs of partial functions.

Before interfacing these ghost objects with concrete C code, we provide a global wrapper allowing us to manipulate the log and table without passing them explicitly as arguments.

\begin{vccdisplay}{Global cryptographic state}
\begin{lstlisting}[language=vcc]
_(ghost _(claimable) struct cryptoState_s
 {
    volatile log L;
    volatile table T;

    _(invariant valid_log(L))
    _(invariant leq_log(\old(L),L))

    _(invariant valid_table(L,T))
    _(invariant leq_table(\old(T),T))
  } CS;)
\end{lstlisting}
\end{vccdisplay}

The two fields are made volatile, and the structure will be initially closed to ensure that all updates have to follow its two-state invariants, stating that the log and table have to grow over all transitions.
Additionally, the one-state invariant states that the log is always valid, ensuring that all the theorems we proved in Coq apply, and that the table is always valid with respect to the log, ensuring that the symbolic security assumptions are never violated.
\end{JCS}

\subsection{The Hybrid Wrappers}\label{sec:hybrid}

We want to ensure that all cryptographic operations are used in ways that preserve the \begin{IEEE}representation table's invariants.\end{IEEE}
\begin{JCS}cryptographic state's invariants.\end{JCS}
We provide hybrid wrappers around the concrete library functions; wrappers are not only verified to maintain the table's invariants but also serve to give symbolic contracts to a cryptographic interface working with concrete bytes.

For simplicity in this paper, the hybrid wrappers manipulate a structure type \kw{bytes_c} containing all information pertaining to a byte array.

\begin{vccdisplay}[-2.5]{A type for byte strings}
\begin{IEEE}
\lstinputlisting[language=VCC,linerange=BytesC,literate={\ bytes_s\ }{\ }{1}]{vcc/v7/hybrids.h}
\end{IEEE}
\begin{JCS}
\begin{lstlisting}[language=VCC,literate={\ bytes_s\ }{\ }{1}]
typedef struct bytes_s {
  unsigned char *ptr;
  unsigned long len;

  _(ghost bytes encoding)
  _(invariant \mine((unsigned char[len]) ptr))
  _(invariant encoding == from_array(ptr,len))
} bytes_c;
\end{lstlisting}
\end{JCS}
\end{vccdisplay}

In particular, we keep not only a pointer to the concrete byte array considered and its length, but we also add a ghost field of type \kw{bytes}, representing---as a mathematical integer---the byte string value contained by the \kw{len} bytes at memory location \kw{ptr}.
For the invariants of \kw{bytes_c} to be admissible, we also make sure, using the \kw{\\mine} keyword, that the heap-allocated byte array of length \kw{len} pointed to by \kw{ptr} is always owned by the structure, ensuring in particular that it is never modified while the structure is kept closed.
\begin{FULL}%
Idiomatic C manipulates the beginning address and length separately.  Porting our method to a lower-level programming style is only a matter of providing the necessary memory-safety annotations, which would in any event be needed to verify non-cryptographic properties of the code, such as memory safety.
\begin{JCS}%
This is discussed further, and applied to some example code, in~\cite{FAST11:Csec}.%
\end{JCS}%
\end{FULL}

As an example, here is the contract of our hybrid wrapper for the \kw{hmac_sha1()} cryptographic function.

\begin{vccdisplay}{Hybrid interface for \kw{hmacsha1()}}
\begin{IEEE}
\lstinputlisting[language=VCC,linerange=hmac]{vcc/v7/hybrids.h}
\end{IEEE}
\begin{JCS}
\begin{lstlisting}[language=vcc]
int hmacsha1(bytes_c *k, bytes_c *b, bytes_c *res
             _(ghost \claim c))
  // Claim property
_(always c, (&CS)->\closed)
  // Properties of input byte strings
_(maintains \wrapped(k))
_(maintains \wrapped(b))
  // Properties of out parameter
_(writes \extent(res), c)
_(ensures !\result ==> \wrapped(res))
  // Cryptographic contract
_(requires CS.T.DefinedB[k->encoding])
_(requires CS.T.DefinedB[b->encoding])
_(ensures !\result ==> CS.T.DefinedB[res->encoding])
    // Cryptographic properties on input terms
_(requires
  canHmac(CS.T.B2T[k->encoding],CS.T.B2T[b->encoding],CS.L)
  || (Level(Low(),CS.T.B2T[k->encoding],CS.L) &&
      Level(Low(),CS.T.B2T[b->encoding],CS.L)))
    // Cryptographic properties on output term
_(ensures !\result ==>
  CS.T.B2T[res->encoding] ==
    Hmac(CS.T.B2T[k->encoding],CS.T.B2T[b->encoding]));
\end{lstlisting}
\end{JCS}
\end{vccdisplay}

\begin{IEEE}
For log and table stability,  as well as concurrent access, there is a ghost parameter containing a claim \kw{c} (represented by the \kw{claimp} macro). The function's  contract says, using the \kw{always(c,...)} construction, that the claim parameter should ensure that the \kw{table} object is closed, and that both table and log have only grown since the claim was created (which is expressed using the \kw{c_stable} variant of the \kw{s_stable} macro). Additionally, the function should ensure that the log and table only grow during its execution.
The next lines of the contract concern memory-safety, e.g., the arguments used to pass values in are \kw{wrapped} both at call-site and at return-site (\kw{maintains(F)} expands to \kw{requires(F) ensures(F)}), and the third argument is used as an output parameter, to return the function's result.
The latter is specified by stating that the function is allowed to \kw{write} the set of memory locations contained within the \kw{span} of the out-parameter. Additionally, we also specify that the embedding (\kw{emb}) of the out-parameter is left \kw{unchanged} by a call to the function, meaning that the pointer refers to the same typed memory location.
It is then specified that a call to this function may write all fields of the structure passed in the \kw{res} argument, and that this memory update has to \kw{wrap} the structure when the function call is successful.
\begin{FULL}%
These are enough to let VCC know that \kw{k} and \kw{b} are not written to during the function's execution.  They are both \kw{wrapped} at call site: their \kw{span} is thus not writable, and they therefore have to be distinct from \kw{res}.%
\end{FULL}

The first three lines under ``Cryptographic contract'' deal only with the table, stating that the input byte strings should appear in the table, and that, upon successful return from the function, the output byte string appears in the table.
Then comes an important cryptographic precondition: that either \textit{MACSays()} holds on the terms associated with the input byte strings (modelling an honest participant's calling conditions), or they are both in \textit{Pub()} (modelling a call by the attacker or a compromised principal).
The postcondition states that, upon successful return, the output byte string is associated with the term obtained by applying the \textbf{Hmac} constructor to the terms associated with the input byte strings.

An honest client, when calling a function with this contract, needs to establish the first disjunct of the precondition: that \textit{MACSays()} holds on the terms associated with the input byte arrays.
For this, to hold, the term associated with the key \kw{k} must be given, in the log, a certain usage $\textbf{HashKey}~hu$ for some hashkey usage $hu$, and the term associated with the message to authenticate \kw{b} must be formatted correctly, as specified by the corresponding inference rule (here \ref{MACSays KeyAB Request} or \ref{MACSays KeyAB Response}).
\end{IEEE}
\begin{JCS}
To ensure that the log and table are kept stable during the call, as well as to allow concurrent updates to the cryptographic states, we pass in, as a ghost parameter, a claim \kw{c} guaranteeing that the global container \kw{CS} remains closed whenever the claim is closed.
All other desired properties (monotonic growth and validity, in particular)
are consequences of this simple fact and can be derived by VCC.
The next lines of the contract deal with memory-safety concerns, in this case expressing the fact that the arguments are wrapped at call-site and return-site, and that the (typed) memory location pointed to by the third argument is written to by the function, and is wrapped on successful return from the function.
The claim \kw{c} is added to the list of objects that may be written by the function in case it is necessary to strengthen the claimed property to help the proof go through.

The first three lines under ``Cryptographic contract'' deal only with the table, stating that the input byte strings should appear in the table, and that, upon successful return from the function, the output byte string appears in the table.
Finally, we require as a precondition that either \coqkw{canHmac} holds on the input parameters in the current cryptographic state (catering for an honest participant's calling conditions), or both the key and payload are \coqkw{Low} (catering for calls by the attacker).

On successful return, we guarantee in the postcondition that the output byte string is associated, in the table, with the term obtained by applying the \coqkw{Hmac} constructor to the terms associated with the input byte strings.

An honest client, when calling this function, will establish that \kw{canHmac} holds on the terms associated with the input byte arrays.
We recall that the definition of \coqkw{canHmac} is in general a disjunction of clauses of the form ``\kw{k} has HMAC key usage \kw{u}, and \kw{m} fulfills the payload condition for \kw{u} in the log''.
In the particular case of our HMAC-based authenticated RPC protocol, an honest participant will know that \kw{u} is indeed \kw{U_KeyAB(a,b)} for some \kw{a} and \kw{b}, and will attempt to prove that the payload is either a well-formatted request on which the \kw{Request} event has been logged for \kw{a} and \kw{b}, or a well-formatted response on which the \kw{Response} event has been logged for \kw{a} and \kw{b}.
\end{JCS}

A typical hybrid wrapper implementation first performs the concrete operation on byte strings (e.g., by calling a cryptographic library) before performing updates on the ghost state to ensure the cryptographic postconditions, whilst maintaining the log and table invariants.
To do so, it first computes the expected cryptographic term by looking up, in the table, the terms associated with the input byte strings and applying the suitable constructor.
Once both the concrete byte string and the corresponding terms are computed, the implementation can check for collisions, and in case there are none, update the table (and the log) as expected. In case a collision happens, an \kw{assume} statement expresses that our symbolic cryptography assumptions have been violated.
\begin{IEEE}\begin{FULL}\pagebreak\end{FULL}\end{IEEE}
\begin{vccdisplay}{A hybrid wrapper for \kw{hmacsha1()}}
\begin{IEEE}
\lstinputlisting[language=VCC,linerange=hmac]{vcc/v7/hybrids.c}
\end{IEEE}
\begin{JCS}
\begin{lstlisting}[language=vcc]
int hmacsha1(bytes_c *k, bytes_c *b, bytes_c *res
             _(ghost \claim c))
{ _(ghost term tb,tk,th)
  _(ghost \bool collision = \false)
  _(assert Level(High(),Hmac(CS.T.B2T[k->encoding],CS.T.B2T[b->encoding]),CS.L))
  _(ghost \claim c0 = \make_claim({ c }, (&CS)->\closed && GrowsCS))

  res->len = 20;
  res->ptr = (unsigned char*) malloc(res->len);
  if (res->ptr == NULL)
    return 1;
  sha1_hmac(k->ptr, _(unchecked)((int) k->len), b->ptr, _(unchecked)((int) b->len), res->ptr);

  _(ghost res->encoding = from_array(res->ptr,res->len))
  _(ghost \wrap((unsigned char[res->len]) res->ptr))
  _(ghost \wrap(res))
  _(ghost _(atomic c0, &CS) {
      tb = CS.T.B2T[b->encoding];
      tk = CS.T.B2T[k->encoding];
      th = Hmac(tk,tb);  // Compute the symbolic term

      if ((CS.T.DefinedB[res->encoding] &&
           CS.T.B2T[res->encoding] != th) ||
          (CS.T.DefinedT[th] &&
           CS.T.T2B[th] != res->encoding))
        collision = \true;
      else
      {
        CS.T.DefinedT[th] = \true;
        CS.T.T2B[th] = res->encoding;
        CS.T.DefinedB[res->encoding] = \true;
        CS.T.B2T[res->encoding] = th;
      }
    })  
  _(assume !collision)  // Our symbolic crypto assumption

  return 0; }
\end{lstlisting}
\end{JCS}
\end{vccdisplay}

Our implementation of an HMAC SHA1 wrapper, shown above, uses the PolarSSL project's \kw{sha1_hmac()} function~(\cite{polarssl}).

The \kw{unchecked} keyword is used to let VCC ignore the potential arithmetic overflow due to the type casts.
\begin{JCS}We could in fact provide annotations that guarantee that the cast never overflows, for example by providing an upper bound on the lengths of keys and bytestrings.\end{JCS}

\begin{JCS}
We also use the \kw{GrowsCS} macro, defined below, which is claimed on the cryptographic state (log and table), guiding VCC through the proof by expanding to the transitive closure of the order on cryptographic states: the cryptographic state at the program point where the claim is created is smaller than the cryptographic state at any program point afterwards.

\begin{vccdisplay}{Transitive closure of the order relation on cryptographic states}
\begin{lstlisting}[language=vcc]
#define GrowsCS\
    (leq_log(\when_claimed(CS.L),CS.L) &&\
     leq_table(\when_claimed(CS.T),CS.T))
\end{lstlisting}
\end{vccdisplay}
\end{JCS}

\begin{IEEE}
Since the table is shared and its fields marked volatile, all reads and writes from and to it need to occur in an atomic block guarded by a claim \kw{c} ensuring, among other things, that the global \kw{table} object is closed.
\end{IEEE}
\begin{JCS}
Since the cryptographic state is shared, and its fields marked volatile, all reads and writes from and to the log and table need to occur in atomic blocks guarded by a claim \kw{c} ensuring, at least, that the global \kw{CS} object is closed.
In our particular case, the claim needs to be slightly stronger, as we rely heavily on the transitivity of the two-state invariant
(an order relation), which VCC does not automatically infer and use.
To strengthen the claim \kw{c}, we simply create a claim \kw{c0} on \kw{c}, whose property immediately follows from the log and table invariants, and guarantees their monotonic growth despite interference from other threads.
\end{JCS}

We also provide a function \kw{toString} (whose contract is shown below), converting an ordinary string pointer to a \kw{bytes_c}, the input type for functions like \kw{hmacsha1}.  It logs a \kw{New} event with usage \kw{AttackerGuess} and assumes the guessed literal does not collide with any other term already in the table.

\noindent\begin{minipage}{\textwidth}
\begin{vccdisplay}{Contract for the \kw{toString} wrapper}
\begin{JCS}%
\begin{lstlisting}[language=vcc]
int toString(unsigned char *in, unsigned long inl, bytes_c *res _(ghost \claim c))
_(maintains \thread_local_array(in,inl) && inl != 0)
_(requires \disjoint(\array_range(in,inl),\extent(res)))
_(writes \extent(res))
_(ensures \result ==> \mutable(res))
_(ensures !\result ==> \wrapped_with_deep_domain(res))
_(always c, (&CS)->\closed)
_(ensures !\result ==> CS.T.DefinedB[res->encoding])
_(ensures !\result ==> Level(Low(),CS.T.B2T[res->encoding],CS.L))
_(ensures !\result ==> res->encoding == \old(from_array(in,inl)))
_(ensures !\result ==> CS.T.B2T[res->encoding] == Literal(res->encoding));
\end{lstlisting}%
\end{JCS}
\end{vccdisplay}
\end{minipage}\smallskip

We also provide a function \kw{bytescmp}, that compares two \kw{bytes_c} objects%
\begin{JCS}, and \kw{pair} and \kw{destruct} functions that marshal and unmarshal \kw{bytes_c} objects into and from injective pairs\end{JCS}.

\subsection{Verifying Protocol Code}

The groundwork needed to specify and verify the RPC protocol code is now complete.
The following shows a slightly simplified version of the annotated code for the client role, where the \textbf{Request} event is logged by the atomic assignment and the final correspondence is asserted as a disjunction of events taking into account the potential compromise of one of the principals involved.
Each of the function calls is verified to happen in a state where the function's precondition holds. In particular, the call to the \kw{channel_write()} function yields a proof obligation that
\begin{IEEE}\textit{Pub()} holds on the term corresponding to the second argument.\end{IEEE}
\begin{JCS}the term corresponding to the second argument is \coqkw{Low} in the current cryptographic state.\end{JCS}
The \kw{return} statements are for various kinds of failure, effectively aborting the client in such cases.
\begin{JCS}
As in the hybrid wrappers, we use a local claim \kw{c0} (with the same claimed property) to encode the fact that the log and table grow with time.
The reference cryptographic state for the transitivity argument is updated, by simply re-claiming the same property in a new state, between all calls to hybrid wrappers.
\end{JCS}

\begin{vccdisplay}{Annotated RPC client code}
\begin{IEEE}
\lstinputlisting[language=VCC,linerange=cameraReady]{vcc/RPC/RPCprot.c}
\end{IEEE}
\begin{JCS}
\begin{lstlisting}[language=vcc]
void client(bytes_c *alice, bytes_c *bob, bytes_c *kab, bytes_c *req, channel* chan _(ghost \claim c))
_(maintains \wrapped(alice) && \wrapped(bob) &&
            \wrapped(kab) && \wrapped(req))
_(always c, (&CS)->\closed)
_(writes c)
_(requires CS.T.DefinedB[alice->encoding] &&
           CS.T.DefinedB[bob->encoding] &&
           CS.T.DefinedB[kab->encoding] &&
           CS.T.DefinedB[req->encoding])
_(requires Level(Low(),CS.T.B2T[alice->encoding],CS.L) &&
           Level(Low(),CS.T.B2T[bob->encoding],CS.L) &&
           Level(Low(),CS.T.B2T[req->encoding],CS.L) &&
           Level(High(),table.B2T[kab->encoding]),CS.L)
_(requires
  RPCKeyAB(CS.T.B2T[alice->encoding],
           CS.T.B2T[bob->encoding],
           CS.T.B2T[kab->encoding],
           CS.L));
{
  _(ghost \claim c0 = createRunningClaim(c))
  bytes_c *toMAC1, *mac1, *msg1;
  bytes_c *msg2, *resp, *toMAC2, *mac2;
  // Event
  _(ghost { _(atomic c, &CS)
    CS.L[Request(CS.T.B2T[a->encoding],
                 CS.T.B2T[b->encoding],
                 CS.T.B2T[req->encoding])] = \true; })
  // Build and send request message
  _(ghost refreshCryptoState(c0))
  if ((toMAC1 = malloc(sizeof(*toMAC1))) == NULL) return;
  if (request(req, toMAC1 _(ghost c))) return;

  _(ghost refreshCryptoState(c0))
  if ((mac1 = malloc(sizeof(*mac1))) == NULL) return;
  if (hmacsha1(kab, toMAC1, mac1 _(ghost c))) return;

  _(ghost refreshCryptoState(c0))
  if ((msg1 = malloc(sizeof(*msg1))) == NULL) return;
  if (pair(req, mac1, msg1 _(ghost c))) return;

  _(ghost refreshCryptoState(c0))
  if (channel_write(chan, msg1 _(ghost c))) return;

  // Receive and check response message
  _(ghost refreshCryptoState(c0))
  if ((msg2 = malloc(sizeof(*msg2))) == NULL) return;
  if (channel_read(chan, msg2 _(ghost c))) return;

  _(ghost refreshCryptoState(c0))
  if ((resp = malloc(sizeof(*resp))) == NULL) return;
  if ((mac2 = malloc(sizeof(*mac2))) == NULL) return;
  if (destruct(msg2, resp, mac2 _(ghost c))) return;

  _(ghost refreshCryptoState(c0))
  if ((toMAC2 = malloc(sizeof(*toMAC2))) == NULL) return;
  if (response(req, resp, toMAC2 _(ghost c))) return;
	
  _(ghost refreshCryptoState(c0))
  if (!hmacsha1Verify(kab, toMAC2, mac2 _(ghost c))) return;

  // Correspondence assertion
  _(assert \active_claim(c0))
  _(assert CS.L[Response(CS.T.B2T[alice->encoding],
                       CS.T.B2T[bob->encoding],
                       CS.T.B2T[req->encoding],
                       CS.T.B2T[resp->encoding])]
        || CS.L[Bad(CS.T.B2T[a->encoding])]
        || CS.L[Bad(CS.T.B2T[b->encoding])]);
}
\end{lstlisting}
\end{JCS}
\end{vccdisplay}

\begin{IEEE}
To prove that the correspondence assertion holds, VCC will use the postconditions of \kw{hmacsha1Verify()} stating that a zero return value implies that the third argument is indeed a valid hmac, the fact that the byte array \kw{toMAC2} is known to have a correct response format as it is the result of a successful call to the \kw{response()} function, and the fact that \textit{Pub()} holds on the response message, as it was read from the network. Using these facts, 
VCC can use the inversion theorems shown in Section~\ref{sec:inductive} and prove the correspondence assertion.
\end{IEEE}
\begin{JCS}
To prove that the correspondence assertion holds, VCC will use the postconditions of \kw{hmacsha1Verify()}, stating that a zero return value implies that either \coqkw{canHmac} holds on the key and payload, or the key used for verifying the MAC is \coqkw{Low}.
This fact, combined with the fact that \kw{toMAC2} is known to have a correct reponse format, as it is the result of a successful call to the \kw{response()} function, and the fact that the response message is \coqkw{Low}, since it is read from the network allows VCC to prove the correspondence assertion by unfolding the definition of \kw{canHmac}.
\end{JCS}

\section{Assumptions Concerning the C Verifier}\label{sec:semantics}

Several research papers \cite{cohen_vcc:practical_2009,cohen_local_2010} document the VCC system but there is no formal model of its semantics of programs and specifications aside from the VCG itself. To be able to formulate a precise specification of the program properties (in particular security properties) verified by VCC, we sketch a conventional operational semantics, in terms of which we specify what we assume about the verifier.   The model sketched here has been formalized as part of our Coq development.
The model idealizes from low-level features of C, using instead a simple Java-like heap model which suffices to formalize the key ideas (just as it does in the paper formalizing the VCC invariant methodology \cite{cohen_local_2010}).
The model serves to define two semantic notions ---assertion failure and safe command--- 
in terms of which we state our assumption that successful verification by VCC implies safety.  
Translating these semantic definitions to an operational semantics of low level C code is routine.
What would be a major undertaking is to formally prove soundness of the VCG.
(Such formalization is reported to be in \correction{progress~\cite{Paul12,CohenPS13}}.)
Our assumption is justified, however, by extensive expert review of the VCG and extensive use of VCC.

\iffull
Leaving aside annotations, a program consists of type and function declarations. The body of a function is a sequence of commands including concurrency primitives: thread fork, send, and receive on named channels (all channels being visible to the attacker).   
Local variables and function parameters (and returns) have declared types and in our model there are no type casts. 
\fi

An \emph{execution environment} consists of a self-contained collection of type and function declarations.
For a given execution environment, a runtime \emph{configuration} takes the form $(h,ts,qs)$ where $h$ is the heap, $ts$ is the thread pool, and $qs$ is a map from channel names to message queues.
A \emph{thread state} consists of a command (its current continuation) and a local \emph{store} (that is, a mapping of locals and parameters to their current values); a \emph{thread poo}l is a finite list of thread states.  
Thus threads share the heap and the message queues (which hold messages sent but not yet received).  
A \emph{run} is a series of configurations that are successors in the transition relation.
The transition relation allows nondeterministic selection of any thread that is not blocked waiting to receive on an empty channel.  A single step (transition) may be an assignment, the test of a branch condition, creation of a new thread, and so forth.  Nondeterministic scheduling models all interleavings including ones that may be preferred by an attacker.

A \emph{state predicate} is a predicate on a heap together with a store.  The store is used for function parameters and results, which are thread local.  The precondition of a function contract is a state predicate; its postcondition is a two-state predicate that refers to the initial and final state of the function's invocation.  An invariant is a predicate on a pointer together with a pair of heaps, as described earlier.

The only unusual feature of the operational semantics is our treatment of assumptions, which are usually only given an axiomatic semantics.  
If there is any thread poised to execute the command \kw{assume}~$p$, and the condition $p$ does not hold in the current configuration, then there is no transition---we say there is an \emph{assumption failure}. If all current assumptions hold, then some thread  takes a step.  Thus some runs end with a ``stuck'' configuration from which there are no successors.  The only other stuck configurations are those where every thread is blocked waiting on an empty channel. 
\iffull
A divergent \kw{atomic} block would also be stuck in our semantics, but we only use manifestly terminating atomic blocks.  We also disallow \kw{assume} statements in atomic blocks. 
\fi

Execution of \kw{assume}~$p$ takes a single step with no effect on state.  
Execution of \kw{assert}~$p$ also has no effect on the state---nor does it have an enabling condition.  An assertion is effectively a labelled skip, in terms of which we formulate correctness.

\begin{definition}[Safe Command]
An \emph{assertion failure} is a run in which there is a configuration where some thread's active command is \kw{assert}~$p$ for some $p$ that does not hold in that configuration, or some object's invariant fails to hold, and there is no assumption failure at that point. 
A configuration is \emph{safe} if none of its runs are assertion failures.  
A command $c$ is \emph{safe} under precondition $p$ if 
for states satisfying $p$, the configuration with that initial state and the single thread $c$ is a safe configuration.
\end{definition}

Given our treatment of assumptions, safety means that there is no assertion failure unless and until there is an assumption failure.

VCC works in a procedure-modular way: it verifies that each function implementation satisfies its contract, under the assumption of specified contracts for all functions directly called in the body.
We formalize this in terms of program fragments, for which 
we use names ending in .c or .h for code or interface texts, as mnemonic for usual file names;
but these may be catenations of multiple files.

\begin{definition}[Verifiable]\label{def:verifJudg}
We write $ api.h\vdash p.c \leadsto q.h  $
to mean there exists $p'.c$ that instruments $p.c$ with additional ghost code (but no \correction{additional} assumptions, and no other changes), and $q'.h$ that may extend $q.h$ with contracts for additional functions (but not alter those in $q.h$) and type invariants, such that VCC successfully verifies the implementation of each function $f$ in $p'.c$ against the contract for $f$ in $q'.h$, under hypotheses $api.h$ and $q'.h$; moreover admissibility holds for all the type invariants.
\end{definition}

\iffull
The most common additions to $p.c$ are assertions that serve as hints to guide the prover, but claims and other ghost code can be added.  Assumptions would subvert the intended specifications and could even be unsound.
The most common additions to $q.h$ are contracts, as $q.h$ may only provide contracts for functions of interest such as \kw{main}, whereas the code $p.c$ may include other functions, which for modular reasoning must have contracts.

There are two interpretations of the phrase ``verified by VCC'' in Definition~\ref{def:verifJudg}.  
The mathematical interpretation is \correction{that} the VCs are valid; this idealizes from limitations of the VCG implementation as well as limitations of the theorem prover used to check validity.  
The pragmatic interpretation is that the actual tool runs successfully (which, modulo bugs, is stronger than the mathematical interpretation).
The results in this section hold for either interpretation.
And of course we have run the actual tool as described in the next sections.  
However, in Section~\ref{sec:security-thm} the key Lemma~\ref{lemma:attack-verifiability-dy} only holds for the mathematical interpretation.
\fi

\iffull
Note that, given headers $p.h$ and $api.h$ and a program $p.c$, VCC never successfully
verifies $p.c$ against $p.h$ with $api.h$ unless $p.c$ compiles against $p.h$ with $api.h$,
which in particular means that $p.h + api.h$ are a closed collection of declarations.
\fi
An immediate consequence of Definition~\ref{def:verifJudg} is the following,
where the $+$ operator stands for catenation.
\begin{lemma}[VCC Modularity]\label{lem:vcc-mod}
If $p.h \vdash q.c \leadsto q.h $ and 
   $p.h+q.h \vdash r.c \leadsto r.h $ then 
   $p.h \vdash q.c+r.c \leadsto q.h+r.h $. 
\end{lemma}


The VCC methodology supports verification conditions for sound modular reasoning, but it is not easy to give a VCG-independent semantics for the verifiability judgement $p.h \vdash q.c \leadsto q.h $.
%
Fortunately, for our purposes we do not need a semantic notion of modular correctness.
It is enough to consider soundness for complete programs.
A complete program is verified as $\emptyset\vdash m.c \leadsto \kw{main.h}$.

\begin{vccdisplay}{\kw{main.h}}
\begin{IEEE}
\lstinputlisting[language=VCC,linerange=Main]{vcc/v7/main.h}
\end{IEEE}
\begin{JCS}
\begin{lstlisting}[language=vcc]
void main()
_(requires \program_entry_point())
_(writes \universe());
\end{lstlisting}
\end{JCS}
\end{vccdisplay}

The
\begin{IEEE}\kw{program_entry_point()}\end{IEEE}
\begin{JCS}\kw{\\program_entry_point()}\end{JCS}
precondition means that all global objects exist and are owned by the current thread at the beginning of this function, as it is the first function that is called when the process is started.
\begin{JCS}
Additionally, the \kw{main} function is allowed to write in the process's entire memory space.
\end{JCS}


\begin{assumption}[VCC Soundness]\label{assume:vcc}
If $\emptyset \vdash m.c \leadsto \kw{main.h}$ then the body of function \kw{main} in $m.c$ is safe for the precondition in \kw{main.h}.   
\end{assumption}

VCC checks that ghost state is used in ways that are sound for reasoning about actual observations; that is, it has no influence on non-ghost state except for introducing additional steps that do not change non-ghost state.
\begin{SHORT}
We formalize this in the long version of the paper.
\end{SHORT}
\begin{FULL}%
\begin{assumption}[VCC Ghost]\label{assume:vcc-ghost}
Let $m.c$ be any complete program, which may include ghost state and ghost code, and let $\widehat{m.c}$ be the program with ghost code (including declarations) removed.
For any run $R$ of $m.c$, let $\hat{R}$ be the sequence of configurations obtained from $R$ by removing ghost code, ghost variables, ghost fields, objects reachable only from ghost variables on the current stack,
and configurations reached by steps of ghost code.  We assume
(a) Ghost code does not introduce new observable behaviours:
For any run $R$ of $m.c$, $\hat{R}$ is a run of $\widehat{m.c}$.
(b) Ghost code does not remove observable behaviours except at assumption failures:
For any run $R$ of $\widehat{m.c}$, there is a run $S$ of $m.c$ such that $R$ is a prefix of $\hat{S}$ and either $R=\hat{S}$ or $R$ ends at an assumption failure.
\end{assumption}%
\end{FULL}

\section{Attack Programs}\label{sec:attack-prog}

An attacker in the symbolic model can intercept messages on unprotected communication links (such as the Internet) and send messages constructed from parts of intercepted messages, as specified by a term algebra.  We model the set of all possible attacks, each attack being represented by an \emph{attack program},
which is C code of a particular form.
In this section, we sketch the formal definition of attack \correction{programs}, relative to a suitable interface, and give an example.
Attack programs are what enable us, in Section~\ref{sec:security-thm}, to use an ordinary program verifier to reason about active attackers.
We considered formalizing standard models of network attackers in terms of our symbolic model, in order to prove that our attack programs do capture the standard notion, but this is straightforward and not illuminating.

An attack program is a straight-line C program that compiles against an \emph{attacker interface}.
Such an interface provides some ``opaque'' type declarations together with some function signatures; these include message send/receive, standard cryptographic operations, and protocol-specific actions like creating sessions and initiating roles.
\begin{FULL}%
For an annotated interface $p.h$, we let $\mathit{erase}(p.h)$ be the attacker interface obtained by deleting annotations and the bodies of type declarations.%
\end{FULL}

\begin{small}
\begin{display}[.3]{Attacker interfaces}
\Category{T}{type}\\
\entry{\kw{bool} \mid \kw{unsigned char*} \mid X\kw{*}}\\
\Category{\mu}{entry in an interface}\\
\entry{\kw{typedef}\ X;}{type declaration}\\
\entry{T\ f(T_1\ x_1, \dots, T_n\ x_n)}{function prototype ($n \geq 0$)}\\ 
\entry{\kw{void}\ f(T_1\ x_1, \dots, T_n\ x_n)}{procedure prototype ($n \geq 0$)}\\
\clause{\iface ::= \mu_1 \dots \mu_n}{interface ($n \geq 0$)}
\end{display}
\end{small}

\begin{SHORT}
For an annotated interface $p.h$, we let $\mathit{erase}(p.h)$ be the attacker interface obtained by deleting annotations and the bodies of type declarations.
\end{SHORT}
Recall the software stack shown in Section~\ref{sec:outline};
the file \kw{RPCshim.h} provides a network attacker interface
including generic cryptography and network operations as well as protocol specific functions.

\begin{vccdisplay}{An attacker interface: $\mathit{erase}$(\kw{RPCshim.h})}
\begin{IEEE}
\lstinputlisting[language=VCC,linerange=EraseAttackInterface]{vcc/RPC/RPCshim.h}
\end{IEEE}
\begin{JCS}
\begin{lstlisting}[language=vcc]
typedef bytespub;

bytespub* att_toBytespub(unsigned char* ptr, unsigned long len);
bytespub* att_pair(bytespub* b1, bytespub* b2);
bytespub* att_fst(bytespub* b);
bytespub* att_snd(bytespub* b);

bytespub* att_hmacsha1(bytespub* k, bytespub* b);
bool att_hmacsha1Verify(bytespub* k, bytespub* b, bytespub* m);

void att_channel_write(channel* chan, bytespub* b);
bytespub* att_channel_read(channel* chan);

typedef session;

session* att_setup(bytespub* cl, bytespub* se);

void att_run_client(session* s, bytespub* request);
void att_run_server(session* s);

bytespub* att_compromise_client(session*s);
bytespub* att_compromise_server(session*s);

channel* att_getChannel_client(session* s);
channel* att_getChannel_server(session* s);
\end{lstlisting}
\end{JCS}
\end{vccdisplay}

Type \kw{bytespub} is critical: its invariant constrains its values to be concrete byte arrays that correspond to
\begin{IEEE}terms that satisfy the \textit{Pub()} predicate.\end{IEEE}
\begin{JCS}\coqkw{Low} terms.\end{JCS}
Verifying the implementation of this attacker interface therefore provides a proof that
\begin{IEEE}\textit{Pub()}\end{IEEE}
\begin{JCS}the set of \coqkw{Low} terms\end{JCS}
is closed under attacker actions.
%
The function contracts in \kw{RPCshim.h} and code in \kw{RPCshim.c} are similar to the hybrid wrappers in Section~\ref{sec:hybrid} but oriented to
\begin{IEEE}Pub\end{IEEE}
\begin{JCS}\coqkw{Low}\end{JCS}
data.  They are more complicated, due to memory safety annotations dealing with thread fork and messaging, though that is mostly protocol-independent. An example contract appears in Section~\ref{sec:security-thm}.

\begin{IEEE}
{\small
\begin{display}{Attack program for given interface $\iface$}
An \emph{attack program} for a given interface $\iface$ has the form:\\
\begin{lstlisting}[language=VCC]
void main() { D C }
\end{lstlisting}
\\where \kw{D} is a sequence of local variable declarations\\
and \kw{C} a sequence of commands, such that:\\
1) Each of the declarations in \kw{D} has the form \kw{T x;}, where \kw{T} is\\
 either \kw{bool}, \kw{unsigned char*}, or \kw{T*} where \kw{T} is declared in $\iface$.\\
2) Each command in the sequence \kw{C} is either
(a) a function call\\assignment with variables as arguments, \kw{x = f(y...);}\\
(b) a procedure call with variables as arguments \kw{f(y...);} \\
or (c) an assignment \kw{x = s;} where \kw{s} is a string literal. \\
3) A variable is assigned at most once and every variable\\mentioned is declared in \kw{D}.\\
4) For each function or procedure call, each argument variable is\\assigned earlier in the sequence of commands.\\
5) In each call to a function or procedure \kw{f}, there is a declaration\\of \kw{f} in $\iface$
and each argument variable in the call has declared type\\identical to that of the corresponding parameter of \kw{f}.\\
6) In a function call assignment \kw{x = f(y...);}, the declared type of\\\kw{x} is the result type of \kw{f}.
In a string assignment \kw{x = s;} the declared\\type of \kw{x} is \kw{unsigned char*}.
\end{display}}
\end{IEEE}
\begin{JCS}
{\small
\begin{display}{Attack program for given interface $\iface$}
An \emph{attack program} for a given interface $\iface$ has the form:\\
\begin{lstlisting}[language=VCC]
void main() { D C }
\end{lstlisting}\\
where \kw{D} is a sequence of local variable declarations
and \kw{C} a sequence of commands,\\such that:\\
1) Each of the declarations in \kw{D} has the form \kw{T x;},\\ where \kw{T} is
 either \kw{bool}, \kw{unsigned char*}, or \kw{T*} where \kw{T} is declared in $\iface$.\\
2) Each command in the sequence \kw{C} is either\\
(a) a function call assignment with variables as arguments, \kw{x = f(y...);}\\
(b) a procedure call with variables as arguments \kw{f(y...);} or\\
(c) an assignment \kw{x = s;} where \kw{s} is a string literal.\\
3) A variable is assigned at most once and every variable mentioned is declared in \kw{D}.\\
4) For each function or procedure call, each argument variable is assigned\\earlier in the sequence of commands.\\
5) In each call to a function or procedure \kw{f}, there is a declaration of \kw{f} in $\iface$\\
and each argument variable in the call has declared type identical to that of\\
the corresponding parameter of \kw{f}.\\
6) In a function call assignment \kw{x = f(y...);}, the declared type of \kw{x} is the result\\type of \kw{f}.
In a string assignment \kw{x = s;} the declared type of \kw{x} is \kw{unsigned char*}.
\end{display}}
\end{JCS}
Owing to item 2, an attack program does not directly assign any object field, nor any global variable. 
Nor does it directly invoke any operations except functions and procedures in $\iface$ (item 5).
We show an 
 example attack program below, that models an attacker that runs an instance of the RPC protocol to completion, passing the messages around correctly.

\noindent\begin{minipage}{\textwidth}
\begin{vccdisplay}{An attack program for \kw{RPCshim.h} (from \kw{RPCattack_0.c})}
\begin{lstlisting}[language=vcc]
void main()
{ unsigned char *a,*b,*r;
  bytespub *alice,*bob,*arg,*req,*resp;
  channel *clientC,*serverC;
  session *s;

  a = "Alice"; alice = att_toBytespub(a,5);
  b = "Bob"; bob = att_toBytespub(b,3);
  r = "Request"; arg = att_toBytespub(r,7);
  s = att_setup(alice,bob);
  clientC = att_getChannel_client(s);
  serverC = att_getChannel_server(s);
  att_run_server(s);
  att_run_client(s,arg);
  req = att_channel_read(clientC);
  att_channel_write(serverC,req);
  resp = att_channel_read(serverC);
  att_channel_write(clientC,resp);}
\end{lstlisting}
\end{vccdisplay}
\end{minipage}\smallskip

\begin{FULL}%
As an extra example, consider the attack program below, which models an attack that would be successful against a 
flawed RPC protocol where the request is not included in the response's MAC:
\begin{display}{A symbolic attack}
A $\rightarrow$ B\=\kill
A $\rightarrow$ B: payload $\vert$ hmac(``request" $\vert$ payload)\\
B $\rightarrow$ A: payload' $\vert$ hmac(``response" $\vert$ payload')
\end{display}

The program would break the correspondence property by making the client accept a response that does not correspond to its request. 

\begin{vccdisplay}{An attack program for \kw{RPCshim.h} (from \kw{RPCattack_1.c})}
\begin{lstlisting}[language=vcc]
void main()
{
  unsigned char *a, *b, *r1, *r2;
  bytespub *alice, *bob, *arg1, *req1, *resp1, *arg2, *req2;
  session *s;
  channel *clientC, *serverC;

  // Setup phase
  a = "Alice"; alice = att_toBytespub(a);
  b = "Bob"; bob = att_toBytespub(b);
  r1 = "Request1"; arg1 = att_toBytespub(r1);
  r2 = "Request2"; arg2 = att_toBytespub(r2);
  s = att_setup(alice, bob);
  clientC = att_getChannel_client(s);
  serverC = att_getChannel_server(s);

  // First run through of the protocol, the attacker observes
  att_run_server(s);
  att_run_client(s, arg1);
  req1 = att_channel_read(clientC);
  att_channel_write(serverC, req1);
  resp1 = att_channel_read(serverC);
  att_channel_write(clientC, resp1);

  // Run only the client, with a different request, and respond with the first run's response
  att_run_client(s, arg2);
  req2 = att_channel_read(clientC);
  att_channel_write(serverC, resp1);
}
\end{lstlisting}
\end{vccdisplay}%
\end{FULL}





\section{An Example Security Theorem}\label{sec:security-thm}

An attack program for the RPC protocol is a program that relies only on \kw{RPCshim.h}.  
To form an executable, it needs to be combined with 
\kw{System} which we define to be the catenation \kw{crypto.c} + \kw{RPChybrids.c} + \kw{RPCprot.c}.  Here \kw{crypto.c} is the library of cryptographic algorithms (and we let it subsume OS libraries, e.g., for memory allocation and sockets), which is used in \kw{RPChybrids.c} and \kw{RPCprot.c}.  

Before providing the formal results, we informally describe a key property on which soundness of our approach rests.  Consider any attack program \kw{M.c} and any run of the program \kw{System + RPCshim.c + M.c}.  It is an invariant that at every step of the run, the representation table holds every term that has arisen by cryptographic computation or by invocation of the \kw{toBytesPub} function, which an attack must use to convert guessed bytestrings to type \kw{bytespub} as needed to invoke the other functions of \kw{RPCshim}.  This is not an invariant that we state in the program annotations;
its only role is to justify our use of assumptions.  The only assumptions used are in \kw{RPCshim.c} and \kw{RPChybrids.c} where collisions are detected.
In light of the key invariant, this means that in any run that reaches an assumption failure, the sequence of terms computed includes a hash collision or an attacker guess of a term that is not public according to the symbolic model of cryptography.  
\iffull
In short, assumptions are used only to \correction{encode the Dolev-Yao model}.
\fi

The contracts in \kw{RPCshim.h} all follow a similar pattern; we give one for reference in the following proof.
\begin{display}{Example contract from \kw{RPCshim.h}}
\begin{IEEE}
\lstinputlisting[language=VCC,linerange=GeneralAttackInterfaceHmac]{vcc/RPC/RPCshim.h}
\end{IEEE}
\begin{JCS}
\begin{lstlisting}[language=vcc]
bytespub* att_hmacsha1(bytespub* k, bytespub* b _(ghost \claim c))
_(maintains \wrapped(k))
_(maintains \wrapped(b))
_(writes k,b,c)
_(always c, (&CS)->\closed)
_(ensures \wrapped(\result));
\end{lstlisting}
\end{JCS}
\end{display}

Attack programs were defined in order to show that their behaviours are among those of interfering threads encompassed by the verification conditions VCC imposes on protocol code.  
By soundness Assumption~\ref{assume:vcc}, this will be a consequence of the following verifiability result.

\begin{lemma}\label{lemma:attack-verifiability-dy}
If $M.c$ is an attack program for  $\mathit{erase}(\kw{RPCshim.h})$, 
then $\kw{RPCshim.h} \vdash M.c \leadsto \kw{main.h}$.
\end{lemma}

As mentioned following Definition~\ref{def:verifJudg}, we cannot prove this result under the ``pragmatic interpretation'' that $\kw{RPCshim.h} \vdash M.c \leadsto \kw{main.h}$ means $M.c$ is verifiable by the VCC tool.  There are infinitely many attack programs, most of which are too large to even fit in storage much less be processed by the tool.    
Instead, we consider the ``mathematical interpretation''; that is, we show that the VCs are valid.
Some parts of the proof can still be done using the tool.

\begin{IEEEproof}
(Sketch)
According to Definition~\ref{def:verifJudg} we have to show admissibility of the type invariants in \kw{RPCshim.h}; this we have checked using VCC.  It remains to prove verifiability of an arbitrary attack program against \kw{main.h}. 

Since the contract in \kw{main.h} does not impose a postcondition and its write specification is vacuous, 
we just need to show that invariants are established and maintained.
Let $M.c$ be \kw{void main()\{D C\}}.
In accord with Definition~\ref{def:verifJudg} we will show verifiability of code \kwCprime\ which augments the statements of \kw{C} with two sorts of instrumentation.  The first is simply to prefix \kw{C} with ghost code that initializes the representation table and log.
\begin{IEEE}
This code is defined as macro \kw{init()}, shown in the code sample below, where maps are defined using VCC's \kw{lambda} notation, and the constants \kw{tagRequest} and \kw{tagResponse} are separately defined to be the integer encoding of the 1-byte-long strings \kw{"1"} and \kw{"2"}, respectively.

\begin{vccdisplay}{The \kw{init()} macro}
\lstinputlisting[language=VCC,linerange=RPCInit]{vcc/RPC/RPCshim.h}
\end{vccdisplay}

We verified a sample attack \correction{program} using \kw{init()}, which serves to prove that \kw{init()} establishes the log and table invariants, as the invariants are proved to hold when the objects are \kw{wrapped}.  Furthermore, \kw{init()} creates a claim \kw{c} on the table (which owns the log) that says they remain wrapped and stable.  Owing to the contracts in \kw{RPCshim.h}, this claim will be maintained, which ensures from that point on that the log and table can never be opened.  So their invariants are maintained even in the presence of interference from interleaved threads.
Thus the second sort of instrumentation in \kwCprime\ passes the claim \kw{c} as ghost parameter to each function and procedure call in \kw{C},
in accord with their contracts in \kw{RPCshim.h}.

\begin{FULL}%
For example: 
\begin{lstlisting}[language=VCC]
  att_run_server(s spec(freshClaim(c)));
\end{lstlisting}%
\end{FULL}
\end{IEEE}
\begin{JCS}%
This code is defined as a ghost function \kw{init()}, whose contract and body are shown in the code sample below, where maps are defined using VCC's \kw{lambda} notation, and the constants \kw{tagRequest()} and \kw{tagResponse()} are separately defined to be the values of type \kw{bytes} representing the 1-byte-long bytestrings \kw{"1"} and \kw{"2"}, respectively.

\begin{vccdisplay}{The \kw{init()} function}
\begin{lstlisting}[language=vcc]
_(ghost void init(out \claim c)
  _(writes \extent(&CS))
  _(ensures \wrapped(&CS))
  _(ensures \wrapped(c) && \active_claim(c))
  _(ensures \claims(c, (&CS)->\closed))
  {
    // Initialize to empty log and table
    CS.L = \lambda event e; \false;
    CS.T.DefinedB = \lambda bytes b; \false;
    CS.T.DefinedT = \lambda term t; \false;

    // Add protocol constants (tags) to log as attacker guesses
    CS.L[New(Literal(tagRequest()),AttackerGuess())] = \true;
    CS.L[New(Literal(tagResponse()),AttackerGuess())] = \true;

    // Add protocol constants (tags) to table as Literalss
    CS.T.DefinedB[tagRequest()] = \true;
    CS.T.B2T[tagRequest()] = Literal(tagRequest());
    CS.T.DefinedT[Literal(tagRequest())] = \true;
    CS.T.T2B[Literal(tagRequest())] = tagRequest();

    CS.T.DefinedB[tagResponse()] = \true;
    CS.T.B2T[tagResponse()] = Literal(tagResponse());
    CS.T.DefinedT[Literal(tagResponse())] = \true;
    CS.T.T2B[Literal(tagResponse())] = tagResponse();

    // Establish invariant and state claim
    \wrap(&CS);
    c = \make_claim({ &CS }, \true);
  })
\end{lstlisting}
\end{vccdisplay}

We used VCC to verify the body of \kw{init()}, which serves as proof that the validity invariants are indeed satisfiable on the initial cryptographic state (composed of the empty log, and the table containing only protocol constants, as attacker guesses).

The function returns, as an out-parameter, a claim \kw{c} on the global cryptographic state, that guarantees that it can never be opened during \correction{execution.  So its} invariants are maintained even in the presence of interference from interleaved threads, which are guaranteed to preserve the two-state invariants through all atomic actions.
Thus, the second sort of instrumentation in \kwCprime\ passes the claim \kw{c} as ghost parameter to each function and procedure call in \kw{C}, in accord with their contracts in \kw{RPCshim.h}.

\begin{FULL}%
For example: 
\begin{lstlisting}[language=VCC]
  att_run_server(s _(ghost c));
\end{lstlisting}%
\end{FULL}%
\end{JCS}

We also add an assertion before each function and procedure call. 
This helps us explain why verification goes through. 
Let us say ``pointer variable'' for the variables declared in \kw{D} with pointer type.  
Preceding each procedure call $f(\overline{y})$ and function call $x = f(\overline{y})$ in \kwCprime\ we can assert a conjunction of the form 
$\kw{wrapped}(x_0) \kw{\&\&} \ldots \kw{\&\&} \kw{wrapped}(x_j)$
where $x_0,\ldots,x_j$ are the pointer variables that have been assigned up to this point.
(We gloss over memory safety assertions needed for bare string literals.)
By induction on the length of \kw{C}, we argue that each of these assertions holds, and moreover the type invariants are maintained.  
%
An assignment, say $x = f(y,z,w);$, satisfies the preconditions of $f$ owing to the added claim, the requirement that $y,z,w$ were previously assigned, and the assertion that $y,z,w$ are all wrapped.
By inspection of the contracts for each $f$ in \kw{RPCshim.h} (e.g., \kw{att_hmacsha1()} given above), that is all that is needed.
The postcondition of $f$ ensures that results are wrapped, so in particular $x$ is wrapped at the next assertion (and the claim maintained).
%
%
\end{IEEEproof}

Running VCC on \kw{RPCattack_0.c} not only served to check the \kw{init()} code used in the proof but also as a sanity check on this Lemma.
In fact, the added intermediate assertions are not needed for VCC to verify the example attack \correction{program}. 



\begin{theorem}\label{thm:secRPC}
Assume $\emptyset \vdash \kw{crypto.c} \leadsto \kw{crypto.h}$.
For any attack program $M.c$ against the interface $\mathit{erase}(\kw{RPCshim.h})$,
the program $\kw{System} + \kw{RPCshim.c} + M.c$ is safe.
\end{theorem}
\begin{IEEEproof}
\ifjcs 
We have verified with VCC that:
\[ \kw{crypto.h} \vdash (\kw{RPChybrids.c}+\kw{RPCprot.c}+\kw{RPCshim.c}) \leadsto \kw{RPCshim.h} \]
By assumption $\emptyset \vdash \kw{crypto.c} \leadsto \kw{crypto.h}$ and Lemma~\ref{lem:vcc-mod} we get
\[ \emptyset\vdash (\kw{crypto.c}+\kw{RPChybrids.c}+\kw{RPCprot.c}+\kw{RPCshim.c}) \leadsto \kw{RPCshim.h} \]
That is, we have
$\emptyset \vdash (\kw{System}+\kw{RPCshim.c}) \leadsto \kw{RPCshim.h}$
by definition of \kw{System}. 
By Lemma~\ref{lemma:attack-verifiability-dy},
since $M.c$ is an attack program for $\mathit{erase}(\kw{RPCshim.h})$, we get
$ \kw{RPCshim.h} \vdash M.c \leadsto \kw{main.h} $.
Hence  by Lemma~\ref{lem:vcc-mod} we get
\[ \emptyset \vdash (\kw{System}+\kw{RPCshim.c} + M.c) \leadsto \kw{main.h} \]
So by Assumption~\ref{assume:vcc} the program $\kw{System}+\kw{RPCshim.c} + M.c$ is safe.
\else 
We have verified with VCC that:\\
{\small$\kw{crypto.h} \vdash (\kw{RPChybrids.c}+\kw{RPCprot.c}+\kw{RPCshim.c}) \leadsto \kw{RPCshim.h}$}
By assumption $\emptyset \vdash \kw{crypto.c} \leadsto \kw{crypto.h}$ and Lemma~\ref{lem:vcc-mod} we get\\
{\footnotesize$\vdash (\kw{crypto.c}+\kw{RPChybrids.c}+\kw{RPCprot.c}+\kw{RPCshim.c}) \leadsto \kw{RPCshim.h}$}
that is, we have
$\emptyset \vdash (\kw{System}+\kw{RPCshim.c}) \leadsto \kw{RPCshim.h}$
by definition of \kw{System}. 
By Lemma~\ref{lemma:attack-verifiability-dy},
since $M.c$ is an attack program for $\mathit{erase}(\kw{RPCshim.h})$, we get
$ \kw{RPCshim.h} \vdash M.c \leadsto \kw{main.h} $
and thus by Lemma~\ref{lem:vcc-mod} we get:
$\emptyset \vdash (\kw{System}+\kw{RPCshim.c} + M.c) \leadsto \kw{main.h}$.
So by Assumption~\ref{assume:vcc} the program $\kw{System}+\kw{RPCshim.c} + M.c$ is safe.
\fi 
\end{IEEEproof}

Informal corollary: For all applications $A$ verified against \kw{RPCprot.h} and the
rest of the API (excluding \kw{RPCshim.h}), $A+\kw{RPCprot.c}+\kw{RPChybrids.c}+\kw{crypto.c}$ is safe in
the presence of any active network attacker (under the symbolic model of cryptography).
The software stack shown in Section~\ref{sec:outline} is executable but its real purpose is to show security for a different software stack, without \kw{RPCshim.c} and \kw{RPCattack\_0.c} but with additional application code that is verified to be memory safe and to conform to the protocol API \kw{RPCprot.h}.

\section{Summary of Empirical Results}\label{sec:additional}

In this section, we summarize our experimental results on implementations of RPC and of the variant of the Otway-Rees protocol presented
by Abadi and Needham~\cite{Abadi:1996:PEP:229713.229714}.

\subsection{Results}
We prove authentication properties of the implementations using non-injective correspondences, expressed as assertions on a log of events, by relying on weak secrecy properties, which we prove formally as invariants of the log.
The attacker controls the network, can instantiate an unbounded number of principals, and can run unbounded instances of each protocol role ---but can never cause a correspondence assertion to fail and can never break the secrecy invariants, unless the Dolev-Yao assumption (no collisions or lucky guesses) has already been violated.
In particular, we prove the following properties about our sample protocol implementations.
\subsubsection{RPC} Our implementation of RPC does not let the server reply to unwanted requests, and does not let the client accept a reply that is not related to a previously sent request. Moreover, their shared key remains secret unless either the client or the server is compromised by the attacker.
\subsubsection{Otway-Rees} The initiator and responder only accept replies from the trusted server that contain a freshly generated key for their specific usage, and this key remains secret unless either the initiator or the responder is compromised.

As both a side-effect and a requirement to use a general purpose verifier, we also prove memory safety properties of our implementations.
This can significantly slow verification, especially in parts of the code that handle the building of messages by catenation, and is a large part of the annotation burden.

\subsection{Performance}
\begin{IEEE}The following table\end{IEEE}
\begin{JCS}Table~\ref{tbl:perf}\end{JCS}
shows verification times, as well as lines of code (LoC) and lines of annotation (LoA) estimations for various implementation files%
\begin{JCS}, and offers a comparison of annotation burden and verification time between the original implementation of the methodology as described in~\cite{DGJN11:Guiding} and the one described in this paper, that leverages more recent VCC features and performance improvements\end{JCS}.
Times are given as over-approximations of the verification time \begin{IEEE}in minutes \end{IEEE}(on a mid-end laptop)%
\begin{JCS}, in seconds\end{JCS}.
The number of lines of annotation includes the function contracts, but not earlier definitions.  For example, when verifying a function in \kw{hybrids.c}, all definitions from \kw{symcrypt.h} can be used but are not counted towards the total.
The shim and sample attack programs are verified, as part of the proof of the security theorem,
but they are not part of the protocol verification and so are omitted here.
\begin{JCS}However, the significant speedup observed on the protocol code is also observable on the shim and attack code.\end{JCS}
\begin{FULL}%
The Otway-Rees shim available online assumes, for simplicity, a special semantics for some function calls, as the threads running the initiator and responder role should be able to return a value to the attacker, which requires some more glue code in C. It is possible to write and verify this glue code using VCC, but it makes the code that much more complex to understand and is not relevant to the protocol's security.%
\end{FULL}

\begin{IEEE}
\begin{footnotesize}
\begin{center}
\begin{tabular}{|l|c|c|c|c|c|}
\hline
File/Function & LoC & LoA & Time (mins)\\
\hline\hline
\kw{symcrypt.h} & - & 50 & $\leq1$\\
\kw{table.h} & - & 50 & $\leq1$\\
\kw{RPCdefs.h} & - & 250 & $\leq1$\\
\kw{ORdefs.h} & - & 250 & $\leq1$\\
\hline\hline
\kw{hybrids.c} & 150 & 300 & $\leq5$\\
\hline
~~\kw{destruct()} & 20 & \textbf{40} & $\leq5$\\
~~\kw{hmacsha1()} & 20 & 20 & $\leq1$\\
\hline\hline
\kw{RPCprot.c} & 130 & 80 & $\leq15$\\
\hline
~~\kw{client()} & 40 & 20 & $\leq5$\\
~~\kw{server()} & 40 & 10 & $\leq10$\\
\hline
\kw{ORprot.c} & 300 & 100 & $\sim 100$\\
\hline
~~\kw{initiator()} & 40 & 15 & $\leq5$\\
~~\kw{responder()} & 100 & \textbf{100} & $\sim 60$\\
~~\kw{server()} & 40 & 15 & $\sim30$\\
\hline
\end{tabular}
\end{center}
\end{footnotesize}
\end{IEEE}
\begin{JCS}
\begin{table*}[h]
\begin{footnotesize}
\begin{center}
\begin{tabular}{|l|c||c|c||c|c|}
\hline
                     &         & \multicolumn{2}{c ||}{Preliminary Version~\cite{DGJN11:Guiding}} & \multicolumn{2}{|c|}{Present Version}\\ \cline{3-6}
File/Function & LoC & LoA & Time (s) & LoA & Time (s)\\
\hline\hline
\kw{symcrypt.h} & - & 50 & $\leq1$ & 5 & -\\
\kw{table.h} & - & 50 & $\leq15$ & 30 & $\leq1$\\
\kw{RPCdefs.h} & - & 250 & $\leq15$ & 200 & $\leq5$\\
\hline\hline
\kw{hybrids.c} & 150 & 300 & $\leq300$ & 200 & $\leq30$\\
\hline
~~\kw{destruct()} & 20 & \textbf{40} & $\leq300$ & 20 & $\leq5$\\
~~\kw{hmacsha1()} & 20 & 20 & $\leq10$ & 20 & $\leq 5$\\
\hline\hline
\kw{RPCprot.c} & 130 & 80 & $\leq900$ & 130 & $\leq60$*\\
\hline
~~\kw{client()} & 40 & 20 & $\leq300$ & 30 & $\leq30$*\\
~~\kw{server()} & 40 & 10 & $\leq600$ & 30 & $\leq30$*\\
\hline\hline
\kw{ORprot.c} & 300 & 100 & $\sim 6000$ & 100 & $\leq200$*\\
\hline
~~\kw{initiator()} & 40 & 15 & $\leq300$ & 18 & $\leq30$*\\
~~\kw{responder()} & 100 & \textbf{100} & $\bot\footnotemark$ & 30 & $\leq140$*\\
~~\kw{server()} & 40 & 15 & $\sim1800$ & 30 & $\leq40$*\\
\hline
\end{tabular}
\caption{Comparison showing changes in number of lines of annotation and improvements in verification times between a preliminary version of our system \cite{DGJN11:Guiding} and the version of this article, for the same C implementations of Authenticated RPC and Otway-Rees.\label{tbl:perf}}
\end{center}
\end{footnotesize}
\end{table*}

The built-in support for induction in VCC, including the stronger type-checking that comes with it, allows the verifier to heavily optimize the background axiomatization for the inductive types.
It also allows a much more succinct and clear definition of the security model, leading to much smaller background axiomatizations.
Additionally, the number of memory-safety-related function contracts has been dramatically reduced by a recent overhaul of the internal axiomatization of the memory model, leading to similar (if not better) verification times with less annotations.
On the other hand, a move by the VCC team to a less \correction{specialised} model of memory allocation (that previously could not interfere with other threads) requires us to keep track much more precisely of the claimed properties on the cryptographic state, increasing the number of (boiler-plate) annotations required to verify the main protocol functions.
Those numbers still remain \correction{on the} order of one line of annotation per line of code.
At the time of writing, the verification times marked with a * \correction{in Table~\ref{tbl:perf}} can only be obtained by passing non-standard options to VCC (present in the source archive's Makefile).
Verification times without this option still show significant improvement over the previous figures.
\footnotetext{This was erroneously reported as a successful verification run in~\cite{DGJN11:Guiding}, due to a VCC bug whereby success was reported when Z3 ran out of memory. The only affected result was a post-condition regarding the fact that the involved principals were distinct, which did not hold when the responder was compromised. In this version, we added a concrete check when the responder receives the initial message, letting her detect the potential attack and close the connection early.}


\end{JCS}

\begin{IEEE}
These two case studies confirm previous observations on the annotation burden that comes with general purpose verifiers (and VCC in particular), in ``order of one line of annotation per line of code'' \cite{cohen_local_2010}, which was also similar in the F7 implementation of the RPC protocol \cite{bhargavan_modular_2010}, where the trusted libraries correspond to our hybrid wrappers.
For the RPC protocol, our verification times are a lot higher than those of F7 (which were all under 30 seconds),
because VCC verifies the program's memory safety simultaneously, whereas F7 relies on F\#'s underlying type system  and memory structure to do so.
\dn{VCC is reasoning about a full program semantics, whereas F7 is checking some lightweight refinement conditions that happen to make sense owing to a bunch of metatheory about the program semantics.  It's not just about memory safety.  But I'm not sure what to say succinctly.}
\fd{I think Andy already mentions this somewhere in the introduction.}
\dn{Yes, fine.  The word ``because'' is too strong here but we can fix this in the long version.}
\FD{This still needs done. how about "in part because ..., and also because VCC is reasoning about a full program semantic, which is abstracted from F7's encoding of refinement types thanks to additional metatheory." [Note that VCC also has some metatheory (subject of sections 2 and 5, I believe), to deal with concurrency]}
\end{IEEE}

So as to focus on verifying security properties, we simplified
\begin{JCS}in the example used\end{JCS}
several aspects of the implementation that were not relevant to symbolic security and usually require extensive annotations: details of network operations are ignored by the verifier (in particular, each principal is only given one single channel to the attacker), and memory is not freed after use.

\begin{JCS}
In a related publication \cite{FAST11:Csec}, we verify a much more idiomatic C program for
\correction{a symmetric-encryption-based RPC protocol (RPC-enc)},
using an axiomatic encoding of usages and logs in VCC, closer to the one presented in~\cite{DGJN11:Guiding}.
The RPC-enc example is not presented in the table above as it was not studied in the preliminary version of our method,
and has not yet been ported over to the current version.
\end{JCS}

\section{Related Work on Protocol Code}
\label{sec:related}

\jj{This is basically the same as in the CSF paper - do we have more space now to include a broader discussion, which may be appropriate for a journal paper ?}
\fd{More worried about time than space at the moment. I'll see what I can do.}

We discussed the closely related tools Csur and ASPIER for C and some tools for {F\#} in Section~\ref{sec:intro}.
We discuss other work on verifying executable code of security protocols.

Pistachio \cite{udrea_rule-based_2006} verifies compliance of a C \correction{program} with a rule-based specification of the communication steps of a protocol.
It proves conformance rather than specific security properties.
DYC \cite{JefLey06} is a C API for symbolic cryptographic protocol messages
which can be used to generate executable protocol implementations,
and also to generate constraints which can be fed to a constraint solver to search for attacks.
Code is checked by model-checking a finite state space rather than being fully verified.

In this paper, we present how a high-level security model can be expressed as part of a C program. Conversely, one can extract a high-level model of the implemented protocol.
Symbolic execution of C code is a promising technique for this purpose.
Corin and Manzano~\cite{CorMan11} extend the KLEE symbolic execution engine to represent the outcome of cryptographic algorithms symbolically,
but do not consider protocol code.
Other recent work \cite{AGJ11} extracts verifiable ProVerif models by symbolic execution of C protocol code, on code similar to that of this paper.
%

There are approaches for verifying implementations of security protocols in other languages.
J{\"{u}}rjens~\cite{jrjens_security_2006} describes a specialist tool to transform a Java program's control-flow graph to a Dolev-Yao formalization in FOL which is verified for security properties with automated theorem provers such as SPASS.
O'Shea \cite{OShea08} translates Java implementations into formal models \correction{in} the LySa process calculus so as to perform a security verification.
The VerifiCard project uses the ESC/Java2 static verifier to check conformance of JavaCard applications to protocol models
(e.g., \cite{HubbersOoostdijkPollSPC03}). 
\begin{FULL}%
Mukhamedov, Gordon, and Ryan \cite{MukGorRya09} perform a formal analysis of the implementation code of a reference implementation of the TPM's authorization and encrypted transport session protocols in F\#, and automatically translate it into executable C code.
\end{FULL}

Work on  RCF, the concurrent lambda calculus underpinning F7, is directly related.
\correction{For example, Backes, Maffei and Unruh~\cite{BMU10:CompSVSC} provide} conditions under which symbolic security of programs in RCF using cryptographic idealizations implies computational security using cryptographic algorithms.
\correction{Also, Backes, Hri\c{t}cu and Maffei~\cite{BacHriMaf11} extend} RCF with union and intersection types for the verification of the source code of cryptographic-protocol implementations in {F\#}.


\begin{JCS}
In recent work, Polikarpova and Moskal~\cite{PM12} show how the cryptographic invariants methodology presented in the conference version of this paper, and improved in this article, can be used to verify security properties of stateful devices.
They modify our approach slightly in two ways.
First, they make use of records and type invariants to define subsets of our \coqkw{Pub} and \coqkw{Bytes} predicates, ensuring only that they are closed under the cryptographic destructors, and modeling attacker actions explicitly, in a way similar to TAPS~\cite{cohen_first-order_2003}.
Second, they rely on an abstract model of the device written and verified in the VCC specification language, and which the C code is proved to refine in VCC.

\correction{More recently, researchers have focused on obtaining computational security properties on implementations using techniques ranging from information-flow analysis~\cite{KustersTG12}, to code extraction~\cite{CadeB13}, to type systems and deductive verification~\cite{BhargavanFKPS13,FournetKS11,Dupressoir2013}, to security-specfic tools and certified compilation~\cite{AlmeidaBBD13}. However, the computational model is outside the scope of the work presented here.}
\end{JCS}

\section{Conclusion}\label{sec:concl}

We describe a method for guiding a general-purpose C verifier to prove both memory safety and authentication and weak secrecy properties of security protocols and their implementations.
Still, our use of VCC leaves clear room for improvement in terms of reducing verification times and \correction{the number} of user-supplied annotations.
Our strategy of building on a general-purpose C verifier aims to benefit from economies of scale,
and in particular to benefit from future improvements in C verification in general.
\begin{IEEE}
This paper establishes a workable method and a baseline.
\end{IEEE}
\begin{JCS}
This paper improves on the baseline performance and generalizes the methods presented in~\cite{DGJN11:Guiding},
and outlines a general way of describing classes of protocols by defining acceptable usages
of cryptographic primitives inductively.
\end{JCS}
We encourage verification specialists to take up the challenge%
\begin{JCS}, as some already did~(\cite{PM12})\end{JCS}.

\begin{IEEE}
We plan, for example, to investigate whether we can improve performance by adopting cryptographic invariants in the style of TAPS~\cite{cohen_first-order_2003}.
Our use of separately-proved secrecy invariants resembles the first-order approach followed in TAPS.
We are aware of unpublished work by Cohen on adapting this approach to use with VCC.
The TAPS style relies less on axioms, which may or may not place more strain on the first-order prover.
\end{IEEE}

Some of our security annotations can be re-used.
In particular, the hybrid wrappers and their contracts need only be written once per cryptographic library, and can be used to verify multiple protocol implementations, as we have done for RPC and Otway-Rees. 
The representation table is also entirely re-usable.
Moreover, we believe that some of the annotations (for example, the log and inductive predicate definitions) may be automatically generated from a high-level description of the protocol.

In future work we intend to adapt our foundations to obtain provably computationally sound results with VCC.
We have designed our contracts to correspond to cryptographic assumptions;
for example, encryptions give only confidentiality and MACs give only integrity.
The table structure and hybrid wrappers introduced in Section~\ref{sec:term-bytes} resemble some standard methods for computational soundness,
such as \correction{the} dual interpretation of the interface in BPW~\cite{backes03composable},
or the hybrid wrappers used in~\cite{Fournet11}.
We may also try more direct methods to obtain computational security results, for example by using the idealized interface from \cite{Fournet11} and by assuming a computationally sound implementation (or linking the C code against the F\# implementation), or by extending the verifier with probabilistic semantics for C, similar to the probabilistic semantics given to the \textsc{pWhile} language in CertiCrypt~\cite{Zanella:2009:POPL}.
\begin{IEEE}
\begin{FULL}
Another needed extension to prove computational security properties would be to verify strong secrecy by observational equivalence.
This would require extending our attacker model slightly to allow attack programs to branch on, for example, MAC verification results.
\end{FULL}
\end{IEEE}

Verification of symbolic security properties remains relevant, even without a computational soundness result, as recent attacks on prominent protocols and implementations could have been found by symbolic protocol verification.
Moreover, some standard features of security protocols (such as sending encrypted keys over the network) are hard to prove secure in the computational model, but may be studied in symbolic models.

For highest assurance, the underlying libraries (\kw{crypto.c}) would be verified, as would any application code using the protocol library.
Moreover, the C verifier would be proved sound with respect to a semantics for which the compiler is proved to be correct
(as in the verified toolchain \cite{Appel11} built on CompCert \cite{DBLP:conf/popl/Leroy06}).
%


\paragraph{Acknowledgements}
Discussions with Misha Aizatulin, Karthik Bhargavan, Josh Berdine, Ernie Cohen, C\'{e}dric Fournet, Bashar Nuseibeh, and Thomas Santen were useful.
Mark Hillebrand and Micha\l\ Moskal helped with VCC methodology, and Stephan Tobies helped with understanding VCC internals.
Naumann acknowledges partial support from Microsoft and NSF awards CCF-0915611 \correction{and CNS-1228930}.

\bibliographystyle{IEEEtranS}

\begin{SHORT}
\input{v7.csf.bbl}
\end{SHORT}
 
\ifdraft
\clearpage
\fi

\begin{JCS}
\appendix
\section{Inductive Definitions for Otway-Rees\label{app:OR}}

As mentioned in the main text, we have also verified a variant of the Otway-Rees key exchange protocol, described in the protocol narration below.
\kw{a} and \kw{b} are the parties establishing a session key, and \kw{s} is a trusted server, assumed to have a shared key \kw{Kp} with every other principal \kw{p}.
\kw{Na} and \kw{Nb} are nonces, although we do not check them for freshness, and \kw{Kab} is a key, freshly generated by the trusted server with the desired usage.

\smallskip
\noindent\begin{minipage}{\textwidth}
\begin{small}
\begin{display}{Additional example: the Otway-Rees key exchange protocol}
a $\rightarrow$ B=\kill
a $\rightarrow$ b: \kw{a | b | Na}\\
b $\rightarrow$ s: \kw{a | b | Na | Nb}\\
s                      \>: \kw{Log(Initiator(a, Na, Kab, b))}\\
s                      \>: \kw{Log(Responder(b, Nb, Kab, a))}\\
s $\rightarrow$ b: \kw{SEnc(Ka, (a | b | Na | Kab)), SEnc(Kb, (a | b | Nb | Kab))}\\
b                      \>: \kw{assert(Responder(b, Nb, Kab, a))}\\
b $\rightarrow$ a: \kw{SEnc(Ka, (a | b | Na | Kab))}\\
a                      \>: \kw{assert(Initiator(a, Na, Kab, b))}
\end{display}
\end{small}
\end{minipage}

This appendix presents the corresponding inductive usage definitions, including payload and compromise predicates.
The usages are defined in such a way that the freshly established shared key can be used to run the Authenticated RPC protocol described in this paper between \kw{a} and \kw{b}.
The full Coq proofs and the verified protocol implementation in VCC are available online (\url{http://fdupress.net/files/journals/guiding/JCS-code.zip}).

\noindent\begin{minipage}{\textwidth}
\begin{vccdisplay}{Coq definitions for terms and usages for Otway-Rees}
\vspace{-1em}
\begin{multicols}{2}
\begin{lstlisting}[language=coq]
Inductive term :=
  | Literal (bs: bytes)
  | Pair (t1 t2: term)
  | Hmac (k m: term)
  | SEnc (k p: term).

Inductive nonceUsage: Type :=.
Inductive hmacUsage: Type :=
  | U_SessionKey (a b: term).
Inductive sencUsage: Type :=
  | U_PrinKey (p: term).
\end{lstlisting}
\columnbreak
\begin{lstlisting}[language=coq]
Inductive usage :=
  | AttackerGuess
  | Nonce (nu: nonceUsage)
  | HmacKey (hu: hmacUsage)
  | SEncKey (su: sencUsage).

Inductive event :=
  | New (l: term) (u: usage)
  | Initiator (p np kpb b: term)
  | Responder (p np kap a: term)
  | Request (a b req: term)
  | Response (a b req resp: term)
  | Bad (a: term).
\end{lstlisting}
\end{multicols}
\vspace{-.5em}
\end{vccdisplay}
\end{minipage}\smallskip

Note that the nonces \kw{Na} and \kw{Nb} are not given a nonce usage as they are sent over the network completely unprotected, and could therefore be replaced with any \coqkw{Low} term by the adversary.

\noindent\begin{minipage}{\textwidth}
\begin{vccdisplay}{Payload and Compromise predicates for Session keys}
\vspace{-1em}
\begin{multicols}{2}
\begin{lstlisting}[language=coq]
Definition SessionKey (a b k: term) (L: log) :=
  Logged (New k (HmacKey (U_SessionKey a b))) L.

Definition SessionKeyComp (a b k: term) (L: log) :=
  Logged (Bad a) L \/
  Logged (Bad b) L.



Definition SessionKeyPayload (a b k m: term) (L: log) :=
  (exists req,
    Requested m req /\
    Logged (Request a b req) L) \/
  (exists req, exists resp,
    Responded m req resp /\
    Logged (Response a b req resp) L).
\end{lstlisting}
\end{multicols}
\vspace{-.5em}
\end{vccdisplay}
\end{minipage}\smallskip

\noindent\begin{minipage}{\textwidth}
\begin{vccdisplay}{Payload and Compromise predicates for Principal Keys}
\vspace{-1em}
\begin{multicols}{2}
\begin{lstlisting}[language=coq]
Definition PrinKey (p k: term) (L: log) :=
  Logged (New k (SEncKey (U_PrinKey p))) L.

Definition PrinKeyComp (p k: term) (L: log) :=
  Logged (Bad p) L.

Definition Pair4 m a b c d :=
  m = Pair a (Pair b (Pair c d)).




Definition PrinKeyPayload (p k m: term) (L: log) :=
  (exists b, exists np, exists kpb,
    p <> b /\
    Pair4 m p b kpb np /\
    SessionKey p b kpb L /\
    Logged (Initiator p np kpb b) L) \/
  (exists a, exists np, exists kap,
    p <> a /\
    Pair4 m a p kap np /\
    SessionKey a p kap L /\
    Logged (Responder p np kap a) L).
\end{lstlisting}
\end{multicols}
\vspace{-.5em}
\end{vccdisplay}
\end{minipage}

\end{JCS}

\begin{IEEE} 
\begin{FULL}
\onecolumn
\appendices 


\section{Files}
For convenience, all VCC and Coq source files listed here, along with the files for the mentioned Otway-Rees example, are made available in a stand-alone archive, available at \url{http://fdupress.net/files/confs/CSF11/guiding-source.tar.gz}.

\subsection{RPCdefs.v with proofs elided (using coqdoc -g)}\label{app:RPCdefs.v}

All the Coq propositions marked as \textbf{Definition}, \textbf{Inductive}, and \textbf{Theorem} are imported into VCC header files,
as described in Section~\ref{sec:symbolic},
but those marked \textbf{Lemma} are not.

\medskip

\begin{small}
\input{Coq/RPCdefs.tex}
\end{small}

\ifdraft
\clearpage
\fi
\subsection{symcrypt.h}\label{app:symcrypt.h}
\lstinputlisting[language=VCC]{vcc/v7/symcrypt.h}

\ifdraft
\clearpage
\fi
\subsection{RPCdefs.h}\label{app:RPCdefs.h}
\hypertarget{app:RPCdefs.h}{}
\lstinputlisting[language=VCC]{vcc/RPC/RPCdefs.h}

\ifdraft
\clearpage
\fi
\subsection{table.h}\label{app:table.h}
\hypertarget{app:table.h}{}
\lstinputlisting[language=VCC]{vcc/v7/table.h}

\ifdraft
\clearpage
\fi
\subsection{hybrids.h}\label{app:hybrids.h}
\hypertarget{app:hybrids.h}{}
\lstinputlisting[language=VCC]{vcc/v7/hybrids.h}

\ifdraft
\clearpage
\subsection{hybrids.c}\label{app:hybrids.c}
\hypertarget{app:hybrids.c}{}
\lstinputlisting[language=VCC]{vcc/v7/hybrids.c}
\else
\subsection{hybrids.c}\label{app:hybrids.c}
\lstinputlisting[language=VCC,linerange=hmac]{vcc/v7/hybrids.c}
\fi

\ifdraft
\clearpage
\fi
\subsection{RPCprot.h}\label{app:RPCprot.h}
\hypertarget{app:RPCprot.h}{}
\lstinputlisting[language=VCC,linerange=forSubmission]{vcc/RPC/RPCprot.h}

\ifdraft
\clearpage
\fi
\subsection{RPCprot.c}\label{app:RPCprot.c}
\lstinputlisting[language=VCC,linerange=forSubmission]{vcc/RPC/RPCprot.c}

\ifdraft
\clearpage
\fi
\subsection{RPCshim.h}\label{app:RPCshim.h}
\hypertarget{app:RPCshim.h}{}
\lstinputlisting[language=VCC]{vcc/RPC/RPCshim.h}

\ifdraft
\clearpage
\subsection{RPCshim.c}\label{app:RPCshim.c}
\lstinputlisting[language=VCC]{vcc/RPC/RPCshim.c}

\clearpage
\subsection{main.h}\label{app:main.h}
\lstinputlisting[language=VCC]{vcc/v7/main.h}
\fi
\end{FULL}
\end{IEEE}

\begin{FULL}
\bibliography{v7}
\end{FULL}

\ifjcs
\end{document}
\fi

\end{document}